\documentclass[pre,aps,showpacs,preprint]{revtex4}

\usepackage{graphicx}
\usepackage{amsfonts}
\usepackage{bm}
\def\be{\begin{equation}}
\def\ee{\end{equation}}
\def\bea{\begin{eqnarray}}
\def\eea{\end{eqnarray}}

\begin{document}

\title{ 
Producing nanodot arrays with improved hexagonal order by patterning surfaces before ion sputtering}

\author{Daniel A. Pearson$^1$, R. Mark Bradley$^1$, Francis C. Motta$^2$, and Patrick D. Shipman$^3$}
\affiliation{[1] Department of Physics, Colorado State University, Fort Collins, CO 80523, USA}
\affiliation{[2] Mathematics Department, Duke University, Durham, NC 27708, USA}
\affiliation{[3] Department of Mathematics, Colorado State University, Fort Collins, CO 80523, USA}
\date{\today}

\begin{abstract}
When the surface of a nominally flat binary material is bombarded with a broad, normally-incident ion beam, disordered hexagonal arrays of nanodots can form. Shipman and Bradley have derived equations of motion that govern the coupled dynamics of the height and composition of such a surface [P.~D.~Shipman and R.~M.~Bradley, Phys.~Rev.~B {\bf 84}, 085420 (2011)]. We investigate the influence of initial conditions on the hexagonal order yielded by integration of those equations of motion. The initial conditions studied are hexagonal and sinusoidal templates, straight scratches and nominally flat surfaces. Our simulations indicate that both kinds of template lead to marked improvements in the hexagonal order if the initial wavelength is approximately equal to or double the linearly selected wavelength.  Scratches enhance the hexagonal order in their vicinity if their width is close to or less than the linearly selected wavelength. Our results suggest that prepatterning a binary material can dramatically increase the hexagonal order achieved at large ion fluences. 
\end{abstract}


\maketitle

\section{Introduction}
\label{sec:intro}

When a nominally flat solid surface is bombarded with a broad ion beam, a variety of self-assembled nanoscale patterns can emerge.  Examples include periodic height modulations or \lq\lq ripples"  \cite{Munoz-Garcia14} as well as nanodots arranged in hexagonal arrays of surprising regularity \cite{Facsko99,Bradley10,Shipman11,Bradley11}.
Ion bombardment therefore has the potential to become a high-throughput, single-step method of mass producing large-area nanostructures with length scales beyond the limits of conventional optical lithography.

The primary obstacle to the widespread adoption of ion-induced pattern formation as a nanoscale fabrication tool has been the presence of numerous defects in the patterns that are typically produced.  In the case of surface ripples, some ripples terminate, while others fuse with their neighbors. In contrast, penta- and hepta-defects are found in hexagonal arrays of nanodots produced by ion bombardment
of binary materials.  

A promising concrete strategy for producing more highly ordered patterns is to bombard a topographically prepatterned surface or \lq\lq template'' rather than an initially flat surface \cite{Cuenat05}.  The template should have a regular structure on a length scale that is longer than the natural spacing of the patterns formed by ion sputtering, so that it can be fabricated by, e.g., optical lithography with a mask or optical standing-wave lithography.  The purpose of the template is to guide the ion-induced self-organization that occurs at shorter length scales, leading to a more highly ordered nanostructure than would be formed on an initially flat surface.

Some steps toward utilizing templates in ion-induced ripple formation on elemental materials have been taken. If a silicon surface is pre-patterned with parallel trenches with a width equal to a few times the ripple wavelength, for example, the ripples that form in the trenches tend to align with the trench walls, and the number of defects in the ripple patterns is small \cite{Cuenat05}. 

An intriguing recent experiment suggests another possible route to enhanced ordering. In the experiment, a silica surface was polished mechanically, producing a set of parallel scratches \cite{Raman12}.  This surface was then subjected to normal-incidence bombardment with a beam of 1.8 MeV gold ions.  The result was an array of nanodots with a much higher degree of order than would have been present had the surface not been polished before bombardment.  In particular, chains of nanodots that were presumably parallel to the scratches were observed. 
Similar results have been obtained if the \lq\lq scratches'' are made by prepatterning the surface using near-grazing-incidence ion bombardment rather than by mechanical polishing \cite{Joe09,Kim11,Kim13}.

In this paper we investigate the efficacy of using a template to improve the order in nanodot arrays produced by normal-incidence ion bombardment of binary materials.  Using numerical simulations, we explore the degree of order produced by a template with a hexagonal array of nanoholes.  The nanohole spacing is chosen to be equal to or longer than the linearly selected wavelength $\lambda_T$, i.e., the natural spacing of the nanodots. We find that this type of template dramatically improves the order when the nanohole spacing is approximately equal to certain integer multiples of $\lambda_T$.  Comparable results are obtained for a template with a sinusoidally varying surface height.  Finally, we study the effect of an initial condition that is meant to resemble a single, long, straight scratch on an otherwise nominally planar surface.  Our simulations show that if the scratch width is appropriately chosen, the degree of hexagonal order is strongly enhanced in its vicinity.

We use three different methods to characterize the degree of hexagonal ordering present in the simulated nanodot arrays. The first is a qualitative method which involves inspection of the peaks in the magnitude of the Fourier transform of the surface height. The other two methods are quantitative. One uses a topological data analysis technique called persistent homology. We will describe how to compute a quantity called the $H_1$ sum, and how it can be used as a sensitive measure of hexagonal order. The second quantitative method involves constructing a Voronoi tessellation for the nanodot peaks and computing its nearest-neighbor number distribution.

The organization of the paper is as follows. In Section~\ref{sec:model}, we briefly introduce the equations of motion that describe the time evolution of the surface of a binary material that is bombarded with a broad, normally-incident ion beam.  These are the equations that we will integrate numerically in our simulations. Section~\ref{sec:methods} describes the numerical method used to solve the equations of motion and the initial conditions we studied. In Section~\ref{sec:methods}, we also discuss how we quantify the hexagonal order in the patterns. Section~\ref{sec:results} contains the results of the numerical simulations for different initial conditions. In Section~\ref{sec:summary}, we discuss the results of our simulations and their implications for future experimental work.

%%%%%%%%%%%%%%%%%%%%%%%%%%%%%%%%%%%%%%
\section{Equations of Motion}
\label{sec:model}

In pioneering work, Shenoy, Chan and Chason studied the coupling between the surface topography and composition that arises during ion bombardment of a binary material \cite{Shenoy07}.  Bradley and Shipman extended this theory to include the effect of mass redistribution \cite{Carter96,Moseler05,Davidovitch07,Madi11,Castro12,Castro12b} and the leading order nonlinear terms for the case of normal-incidence ion bombardment \cite{Bradley10,Shipman11,Bradley11}.  The Bradley-Shipman equations of motion govern the behavior of $u$ and $\phi$, the deviations of the surface height and surface composition from their steady-state values.  Adopting the same notation and assumptions as Bradley and Shipman, the equations are
\be
\label{eq:eom1}
\frac{\partial u}{\partial t} = \phi - \nabla^2 u -\nabla^2\nabla^2 u + \lambda\left(\nabla u\right)^2
\ee
and
\be
\label{eq:eom2}
\frac{\partial \phi}{\partial t} = -a\phi + b\nabla^2 u +c\nabla^2\phi+\nu\phi^2+\eta\phi^3,
\ee
where the variables $x$, $y$, $t$ and $u$ have been rescaled so that they are dimensionless.
The coefficients $a,b,c,\lambda,\nu$ and $\eta$ depend on the choice of binary material and ion beam. Explicit formulae that relate these coefficients to the underlying physical parameters may be found in Ref.~\cite{Bradley11}.  A discussion of the physical meaning of all of the terms in the equations of motion may also be found there.

%%%%%%%%%%%%%%%%%%%%%%%%%%%%%%%%%%%%%%%
\section{Numerical Methods}
\label{sec:methods}
In order to reduce boundary effects in our numerical integrations of the equations of motion (\ref{eq:eom1}) and (\ref{eq:eom2}), we adopted periodic boundary conditions. We used a Fourier spectral method on a grid of $256\times 256$ points in our integrations of the equations of motion. The linear parts of the equations of motions were integrated in Fourier space, and the nonlinear parts were evaluated in real space. Time stepping was carried out using a fast method introduced by Cox and Matthews: fourth-order Runge-Kutta exponential time-differencing \cite{Cox02,Kassam}. In all of the simulations, the parameter values $a = 0.25$, $b = 0.37$, $c = 1$, $\eta=10$ and $\lambda=0$ were used. This choice of the coefficients $a$, $b$ and $c$ guarantees that there is a narrow band of unstable wave numbers, which is necessary for hexagonal ordering \cite{Bradley10,Shipman11,Bradley11}. For these parameter values, the wavelength with the highest linear growth rate is $\lambda_T\simeq 10.26$. The value of the parameter $\nu$ determines whether the long-time pattern consists of stripes or nanodots \cite{Bradley10,Shipman11,Bradley11}. Since stripes have not been observed in experimental studies of normal-incidence bombardment of binary materials, we used $\nu=1$ in all simulations, and so obtained patterns composed of nanodots.

\subsection{Initial Conditions}
Three types of initial conditions will be considered in this paper: hexagonally ordered arrays of nanoholes, sinusoidal ripples and straight scratches. We superimposed small amplitude spatial white noise on the initial conditions to account for the randomness which would exist on a real prepatterned surface. The initial condition for the composition was small amplitude spatial white noise in all simulations. The noise had a maximum amplitude of $10^{-4}$ for both the height and composition in all of the simulations. 

The hexagonal initial condition was formed by superimposing three sine waves. In order to satisfy the periodic boundary conditions, we chose to do the simulations in the rectangular domain given by $-L\le x\le L$ and $-L/\sqrt{3}\le y \le L/\sqrt{3}$, where $L=200$. The functional form used for the hexagonal initial condition was
\be
\label{eq:hex_init}
u_{\mathrm{hex},0}(x,y) = 10^{-2}\left[\sin^2{(\mathbf{k}_\mathrm{a}\cdot\mathbf{r})}
+\sin^2{(\mathbf{k}_\mathrm{b}\cdot\mathbf{r})}
+\sin^2{(\mathbf{k}_\mathrm{c}\cdot\mathbf{r})}\right]+\eta(x,y),
\ee  
where $\eta(x,y)$ is the low amplitude spatial white noise, $\mathbf{r}\equiv x\hat{\mathbf{x}}+y\hat{\mathbf{y}}$, $\hat{\mathbf{k}}_\mathrm{a} \equiv \hat{\mathbf{x}}$, $\hat{\mathbf{k}}_\mathrm{b} \equiv \cos{(2\pi/3)}\hat{\mathbf{x}}+\sin{(2\pi/3)}\hat{\mathbf{y}}$, $\hat{\mathbf{k}}_\mathrm{c} \equiv \cos{(4\pi/3)}\hat{\mathbf{x}}+\sin{(4\pi/3)}\hat{\mathbf{y}}$, and $k_\mathrm{a}$, $k_\mathrm{b}$ and $k_\mathrm{c}$ are set to a common value which we will call $k_I$. Since each of the sinusoids is squared, $\lambda_1\equiv \pi/k_I$ is their wavelength. We varied the parameter $\lambda_1$ from simulation to simulation while keeping $L$ fixed. For convenience, let $k_1 \equiv 2\pi/\lambda_1 = 2k_I$. The wavelength $\lambda_1$ cannot be chosen arbitrarily, since $2L/\lambda_1$ must be a positive integer. If $2L/\lambda_1$ were not an integer, then, because of the periodic boundary conditions, there would be an unphysical discontinuity in the height profile of the initial condition, which would produce unphysical results.

The functional form for the sinusoidal initial condition was 
\be
\label{eq:sin_init}
u_{\mathrm{sin},0}(x,y) = 10^{-2}\sin{\left(k_2 x\right)}+\eta(x,y),
\ee
where $k_2\equiv2\pi/\lambda_2$ and $\lambda_2$ is the wavelength of the initial sinusoid, which we varied between different simulations. The spatial domain was taken to be square: $-L\le x,y\le L$. Again, the initial wavelength could not be chosen arbitrarily: $2L/\lambda_2$ must equal a positive integer. 

The functional form for the scratch initial condition was motivated by an experiment in which an atomic force microscope was used to scratch a Ni-Fe surface \cite{Tseng11}. The scratching process produced ridges on each side of the groove --- a feature which we included in our initial condition. The form of the initial condition used for the scratch template was
\be
\label{eq:scratch_init}
u_{\mathrm{scratch},0}(x,y) = 10^{-2}\left(\frac{x^2}{1.25\sigma^2}-1\right)\exp{\left(-\frac{x^2}{2\sigma^2}\right)}+\eta(x,y),
\ee
where $\sigma$ is a parameter that determines the half-width of the scratch. Varying $\sigma$ does not affect the scratch's maximum or minimum values. The number $1.25$ appears only in order to produce a reasonable ridge-height-to-scratch-depth ratio. Since we are using periodic boundary conditions, this single scratch on a finite spatial domain can be thought of as a series of widely-spaced, parallel scratches on an infinitely extended domain. The scratches lie parallel to the $y$-axis.

\subsection{Quantifying Order}
\label{quantifying_order}
\subsubsection{Fourier Space}
\label{ft_order}
The degree of ordering of the surface may be seen qualitatively by looking at the Fourier transform of the surface height. For example, if the system forms a hexagonally ordered array of nanodots, then a well-ordered pattern will exhibit six strong peaks in Fourier space: The peaks will be located near the circle in $\bm{k}$-space given by $k_x^2+k_y^2=k_T^2$ (where $k_T\equiv 2\pi/\lambda_T$) and will be separated by an angle of $60^\mathrm{o}$. On the other hand, a disordered pattern of nanodots will not exhibit strong peaks in Fourier space.  

\subsubsection{Persistent Homology}
\label{ph_order}
In this subsection, we describe a method of quantifying hexagonal order that is based on a topological data analysis technique known as persistent homology \cite{survey}. A brief overview of our method is given in this subsection; for details and the larger mathematical context, see Ref.~\cite{ph_motta_shipman}.  

\begin{figure}[htp]
\centering
\begin{tabular}{@{}ccc@{}}
\includegraphics[width=0.325\textwidth]{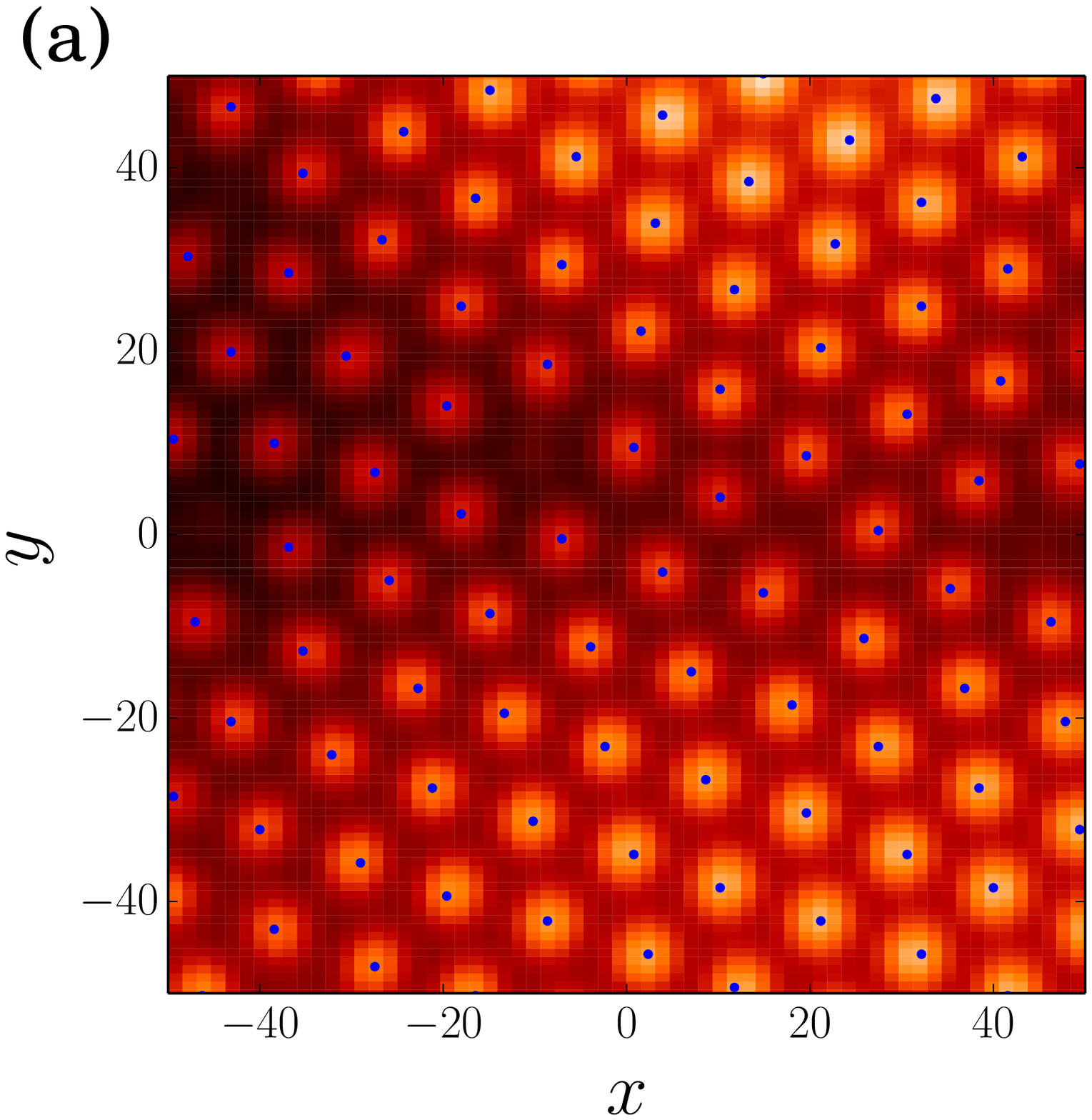} &
\includegraphics[width=0.325\textwidth]{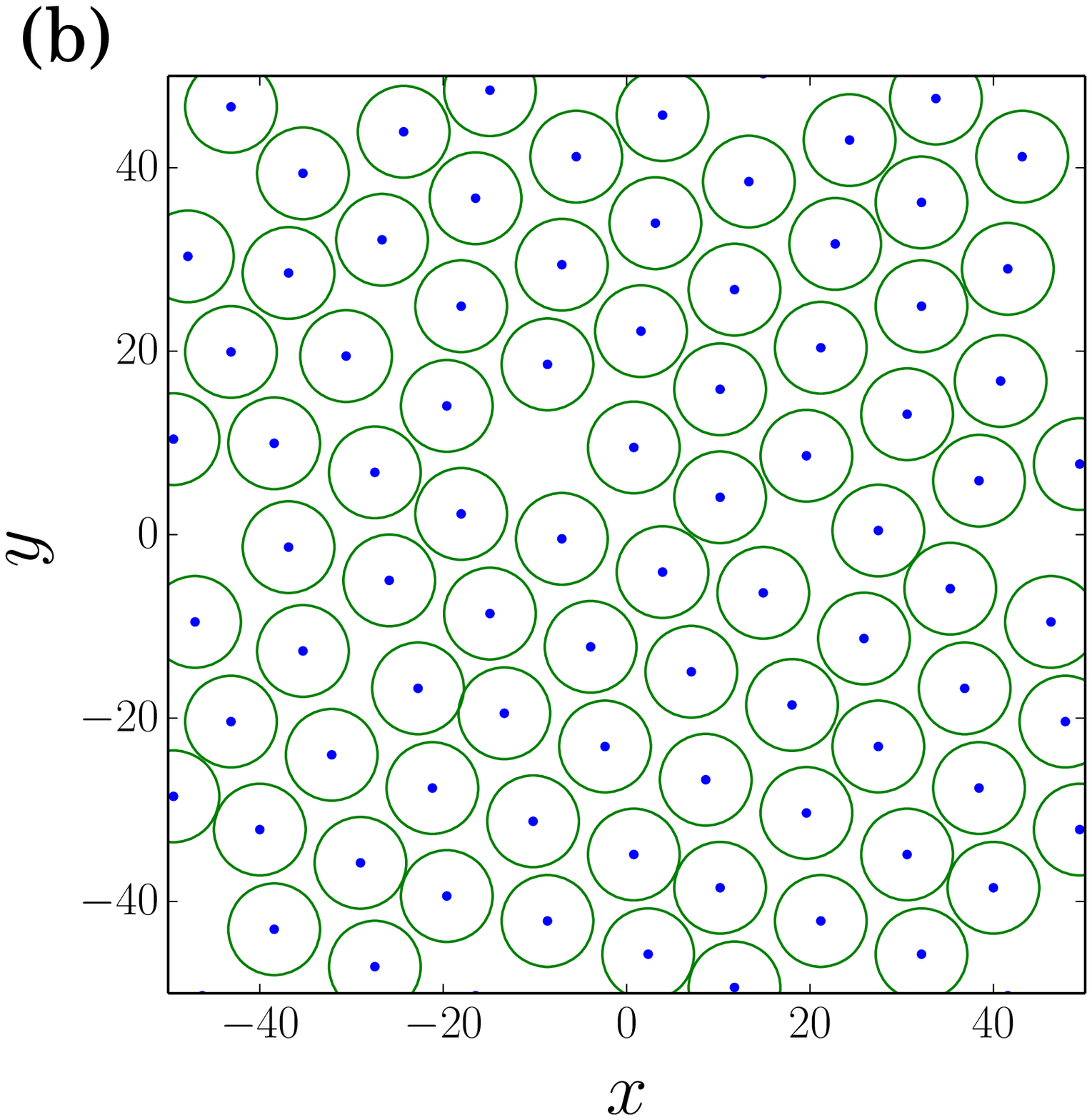} &
\includegraphics[width=0.325\textwidth]{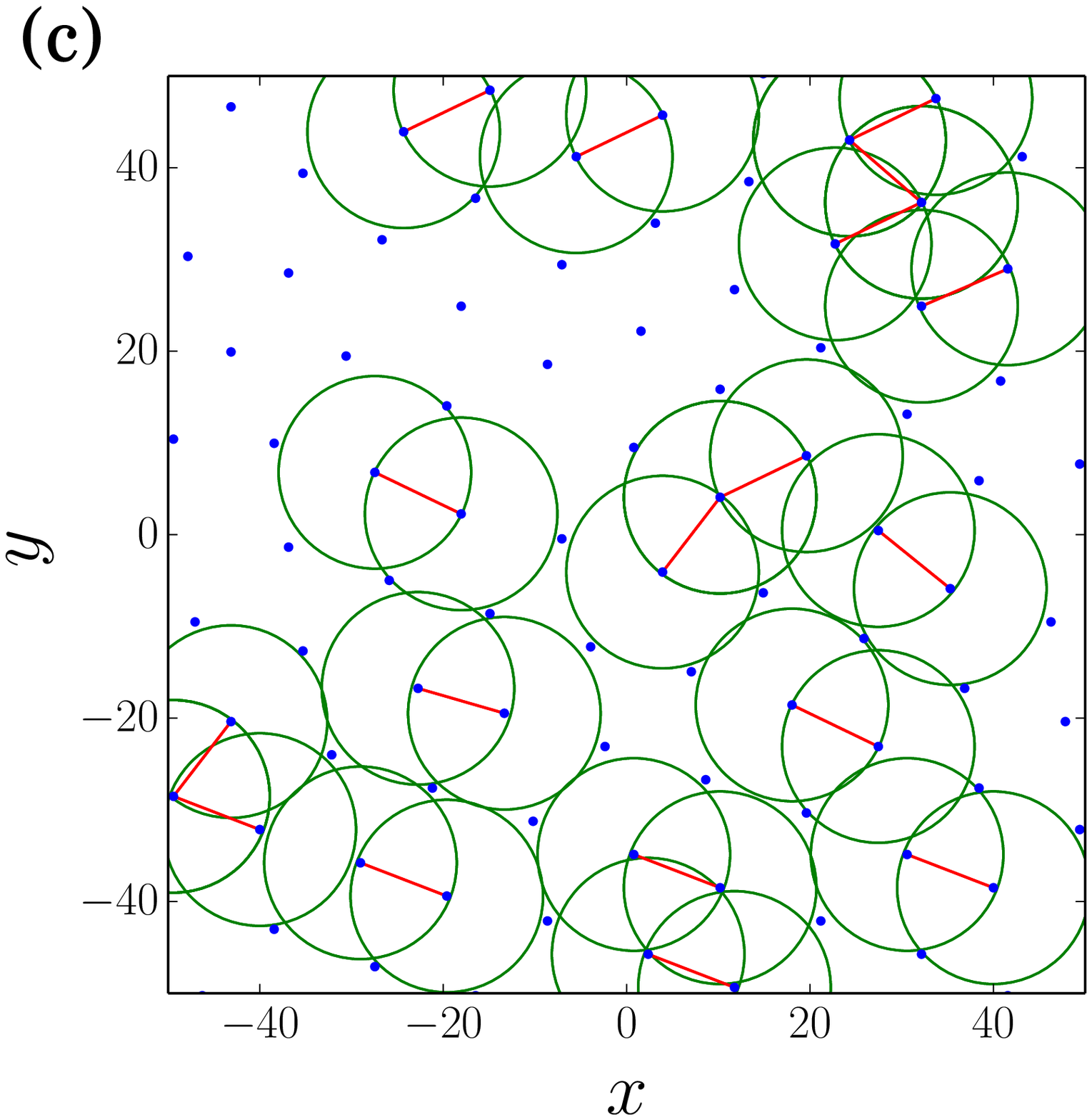} \\
\includegraphics[width=0.325\textwidth]{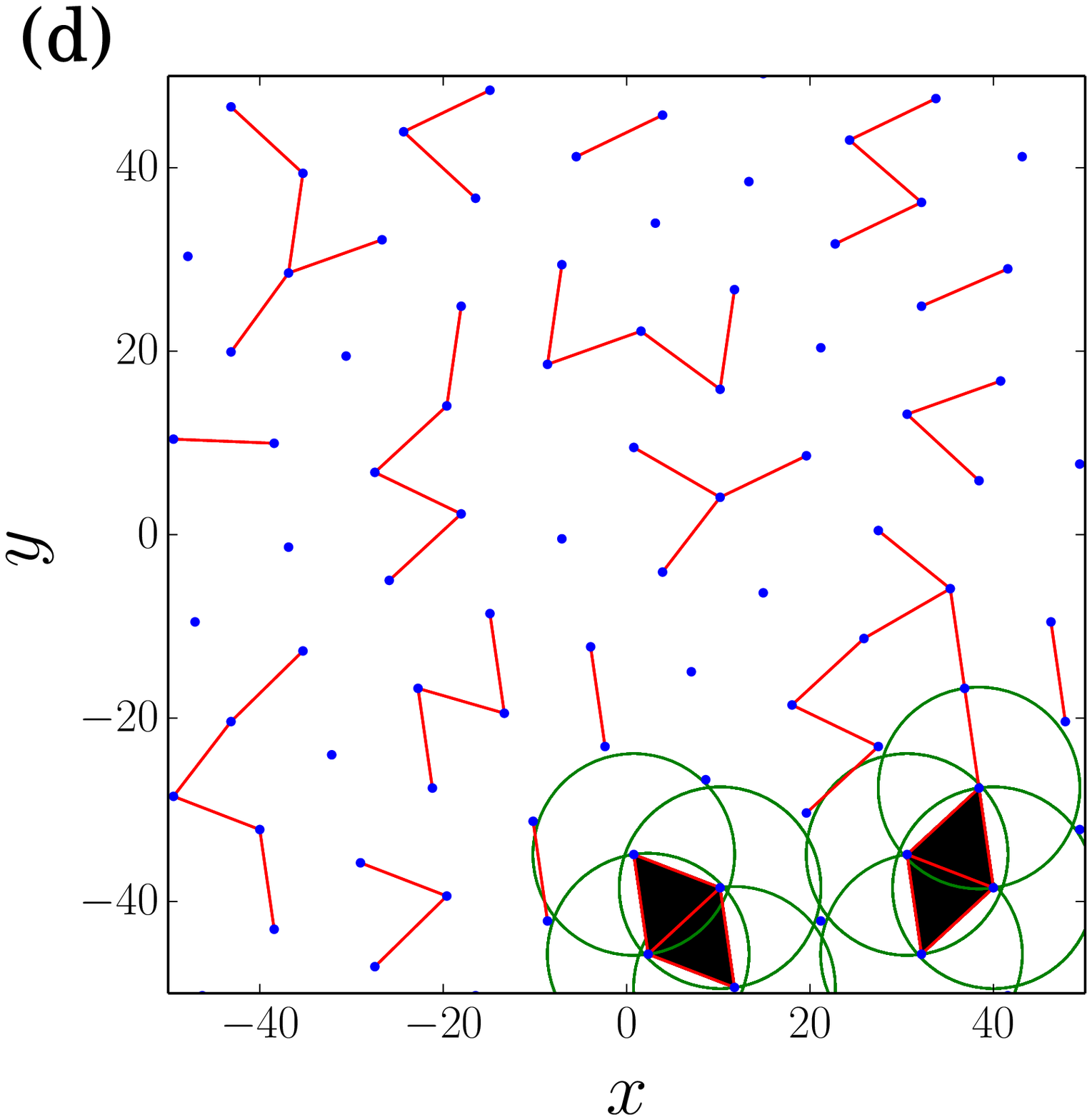} &
\includegraphics[width=0.325\textwidth]{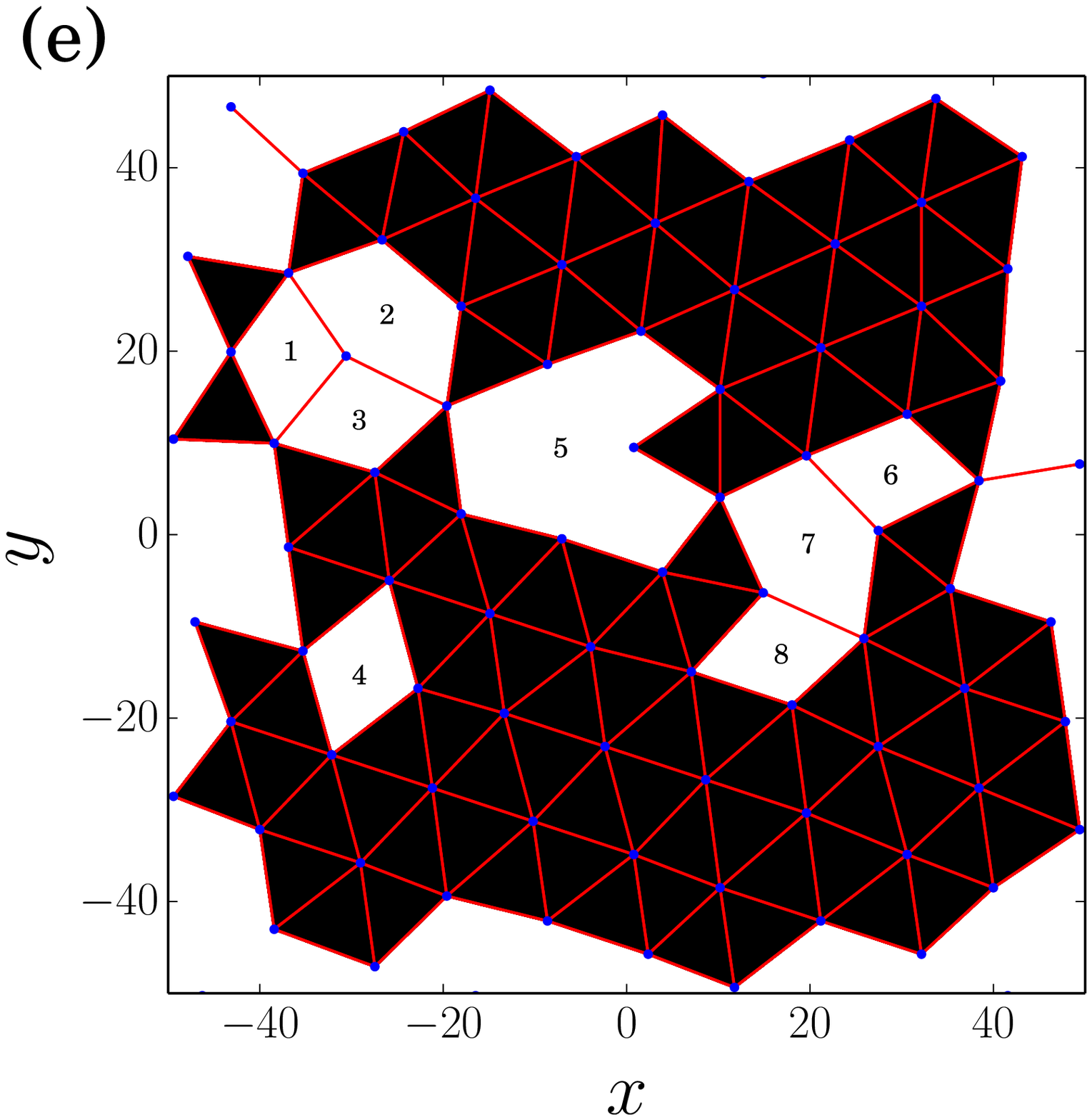} &
\includegraphics[width=0.325\textwidth]{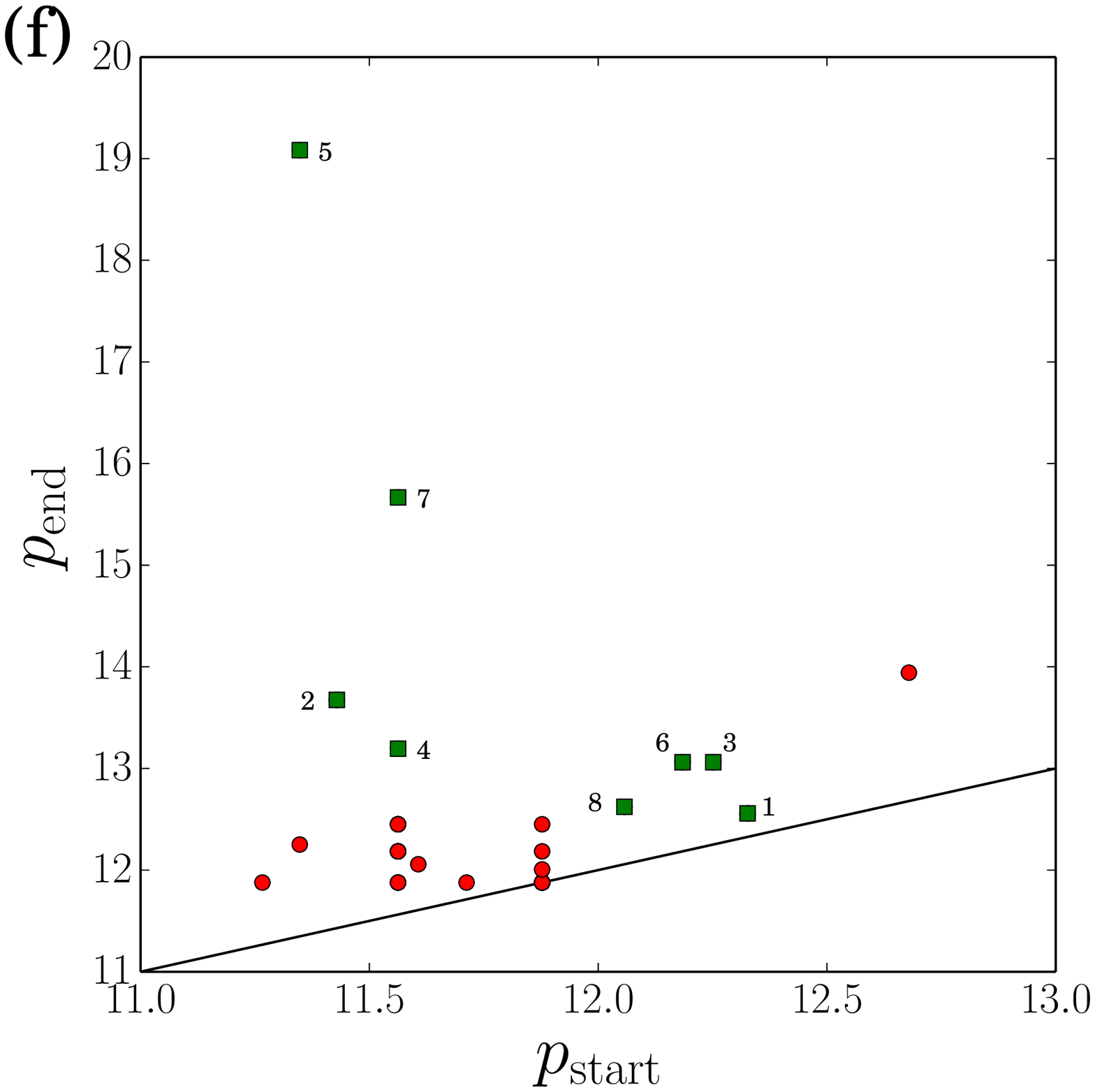}
\end{tabular}
		\caption{(color online) \textbf{(a)} A section of the simulation result shown in Fig.~\ref{fig:no_templating_dots_rect} with blue dots indicating the nanodot peaks. \textbf{(b)} Around each peak, a circle of radius $p=5$ has been drawn. \textbf{(c)} The connectivity parameter $p$ has increased to 10.5 and red edges have been placed between the centers of circles that enclose each other's centers. Only the circles that enclose each other's centers are shown for clarity. \textbf{(d)} When $p=11$, there are four filled in triangles where three circles all enclose each other's centers. Only the circles related to the filled in faces are shown. \textbf{(e)} When $p=12.5$, many faces have been filled in and holes have emerged. The 8 white polygonal regions with red edge boundaries are identified as holes at this value of $p$. \textbf{(f)} A plot of $p_\mathrm{end}$ versus $p_\mathrm{start}$ for the holes found by our persistent homology analysis of (a). The green squares (red circles) are holes which are present (absent) in (e).  The line $p_\mathrm{end}=p_\mathrm{start}$ is also shown. A point's vertical distance above the line is the corresponding hole's persistence interval length.}
\label{fig:ph}
\end{figure}

To quantify the hexagonal order in a pattern of nanodots, we start by obtaining a discrete set of points from a surface pattern by recording the $(x,y)$ coordinates of each nanodot peak, as in Fig.~\ref{fig:ph}(a). As in Fig.~\ref{fig:ph}(b), a circle of radius $p$ is drawn around each of these points. The radius $p$ is called the connectivity parameter; it will be increased from 0 to some maximum value. Clearly, for sufficiently large $p$, some of the circles will enclose each other's centers. For every two circles that enclose each other's centers, we connect the corresponding center points by an edge, as shown in Fig.~\ref{fig:ph}(c). Every time three circles enclose each other's centers, we fill in the triangle which has the centers of the circles as its vertices (Fig.~\ref{fig:ph}(d)), yielding a face. Finally, for a given value of $p$, a hole is identified whenever edges form the boundary of an unfilled region. For example, Fig.~\ref{fig:ph}(e) shows 8 holes for $p=12.5$. Note that hole $\# 5$ corresponds to the largest defect seen in Fig.~\ref{fig:ph}(a). 

A hole's persistence interval length equals $p_\mathrm{end}-p_\mathrm{start}$, where $p_\mathrm{start}$ is the $p$ value at which the hole forms, and $p_\mathrm{end}$ is the $p$ value at which the hole gets filled in and ceases to exist. Summing up the lengths of all the persistence intervals gives a nonnegative number which we will call the $H_1$ sum. 

There are multiple open-source software packages capable of computing persistence intervals for a set of discrete points.  We used the R package called {\it phom} in our analysis~\cite{phom}.

In Fig.~\ref{fig:ph}(f), each hole {\it phom} identified in Fig.~\ref{fig:ph}(a) is represented by a point.  The coordinates of a point are the values of $p_\mathrm{start}$ and $p_\mathrm{end}$ for the hole in question.  The persistence interval lengths are the vertical distances of the points above the line given by
$p_\mathrm{start}=p_\mathrm{start}$.
The 8 green squares are those with $p_\mathrm{start} \leq 12.5 \leq p_\mathrm{end}$ and so can be seen in Fig.~\ref{fig:ph}(e); the red circles are holes not seen in Fig.~\ref{fig:ph}(e).

Since a perfect hexagonal array of points is composed of equilateral triangles, every time three edges form a triangle, the corresponding three circles will enclose each other's centers, and therefore every triangle will be filled in at the same value of $p$. Thus, a persistent homology computation of a perfectly ordered hexagonal array of points will find no holes for any value of $p$, and so the $H_1$ sum will be zero. Whenever there is a vacancy or another type of disorder in the hexagonal lattice, the persistent homology analysis will reveal the presence of one or more holes. Therefore, we can quantify the amount of disorder in an imperfect hexagonal array of points using the $H_1$ sum: The smaller the $H_1$ sum, the better the hexagonal order.

\subsubsection{Nearest-Neighbor Distribution}
\label{nn_order}
A second approach we will use to quantify hexagonal order involves computing the number of nearest neighbors each nanodot has. 
As in the persistent homology approach, we first identify the nanodot peaks and obtain the $(x,y)$ coordinates of these peaks. Let $m$ be the number of points obtained and call the points $\bm{q}_i$ for $i=1,...,m$. From this set of points, we construct the Voronoi tessellation  --- this partitions the plane in way such that each $\bm{q}_i$ lies within the polygon consisting of all points closer to $\bm{q}_i$ than to any $\bm{q}_j$ with $j \neq i$.  The number of polygons in the tessellation with $n$ sides will be denoted by $\Lambda(n)$.  Thus, $\Lambda(n)$ is the number of points $\bm{q}_i$ with $n$ nearest neighbors and will be referred to as the nearest-neighbor distribution.  For a perfectly hexagonal lattice, the mean and variance of the nearest-neighbor distribution are exactly six and zero, respectively. Thus, to quantify hexagonal order, we will compute the mean and variance of $\Lambda(n)$ and compare them to these numbers.
%%%%%%%%%%%%%%%%%%%%%%%%%%%%%%%
\section{Results}
\label{sec:results}
We separate our simulation results into four subsections: (A) nominally flat initial surfaces, (B) hexagonal templates, (C) sinusoidal templates and (D) scratch initial conditions.
\subsection{Nominally Flat Initial Conditions}
\begin{figure}[htp]
\centering
\begin{tabular}{@{}cc@{}}
\includegraphics[width=0.45\textwidth]{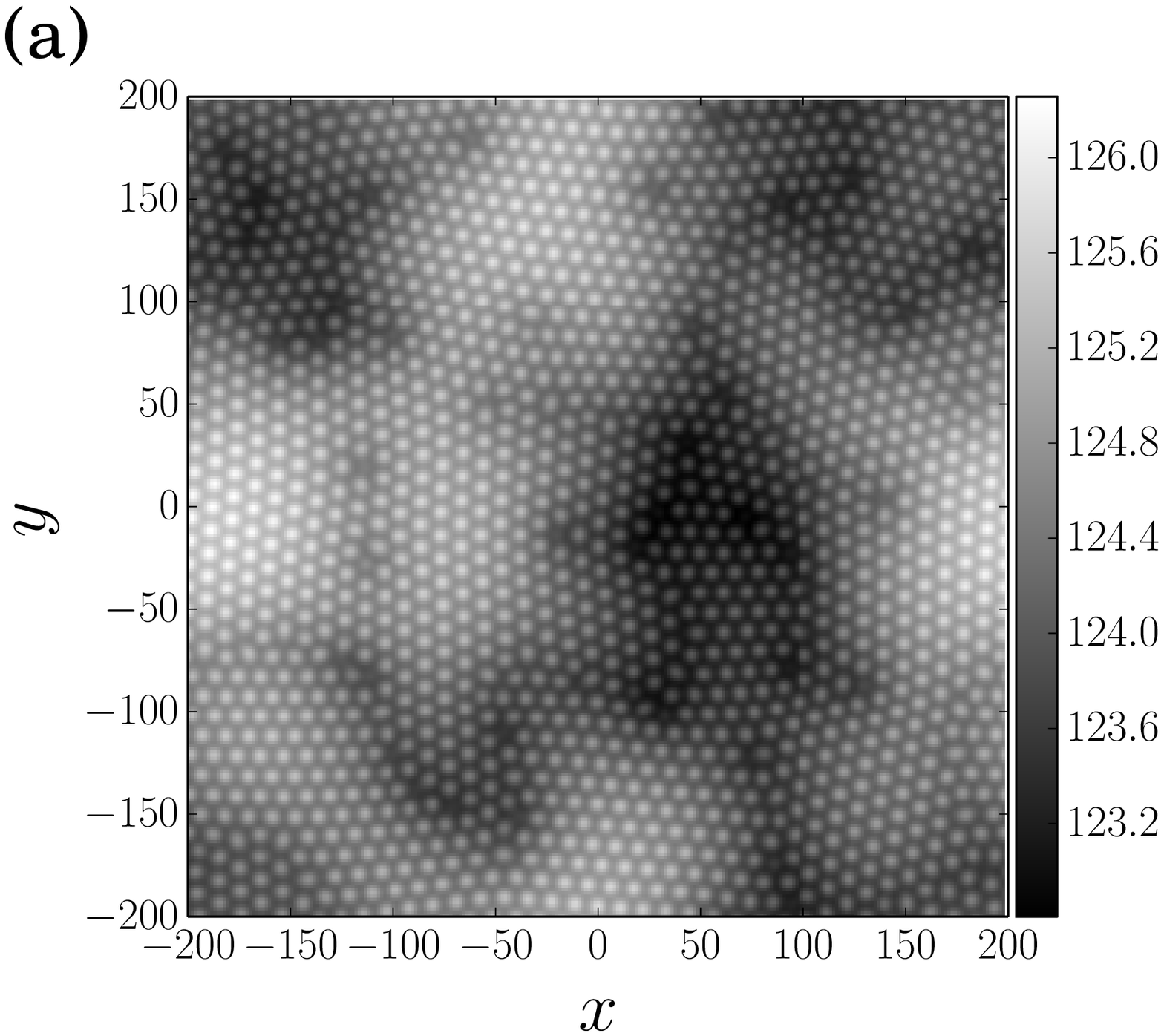} &
\includegraphics[width=0.45\textwidth]{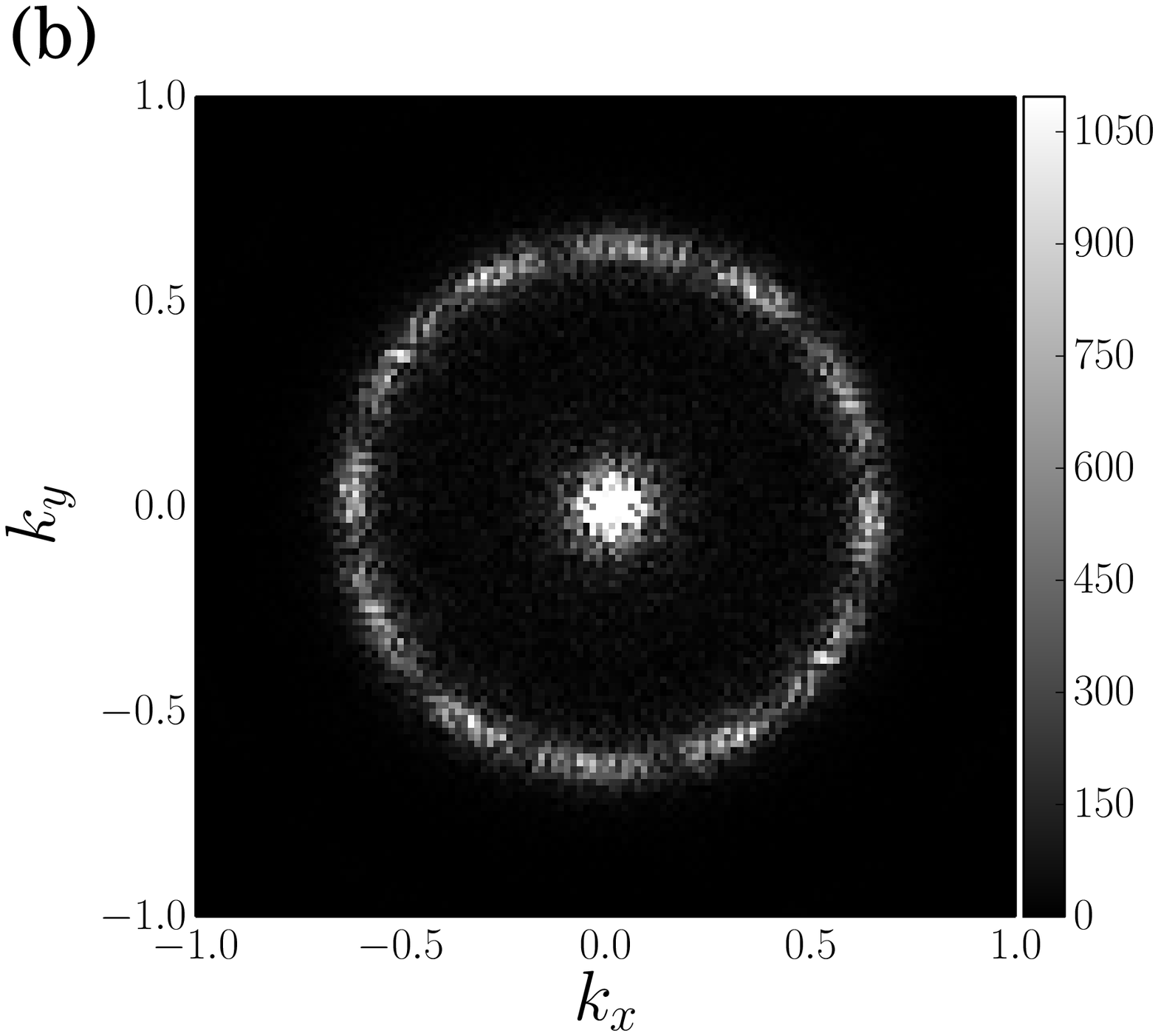}
\end{tabular}
\caption{The height (a) and the magnitude of the Fourier transform (b) of a non-templated surface after integrating to time $t=10^4$. In order to prevent the central peak from dominating the plot, the gray scale in (b) is capped at the maximum value of the magnitudes of the Fourier modes within the annulus of linearly unstable wave vectors.  This is also done in all subsequent figures showing Fourier transforms. }
\label{fig:no_templating_dots}
\end{figure}

\begin{figure}[htp]
\centering
\begin{tabular}{@{}cc@{}}
\includegraphics[width=0.45\textwidth]{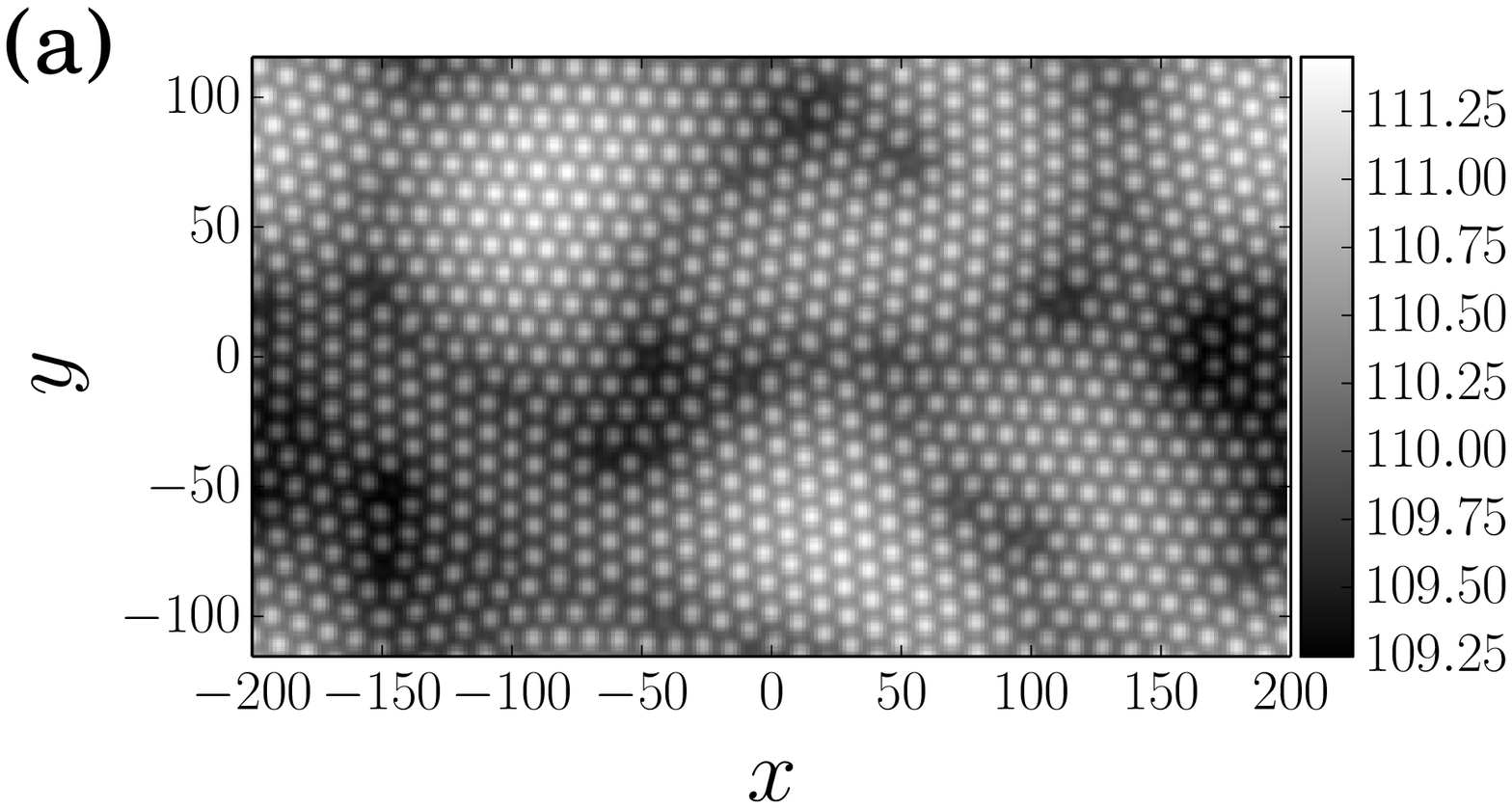} &
\includegraphics[width=0.45\textwidth]{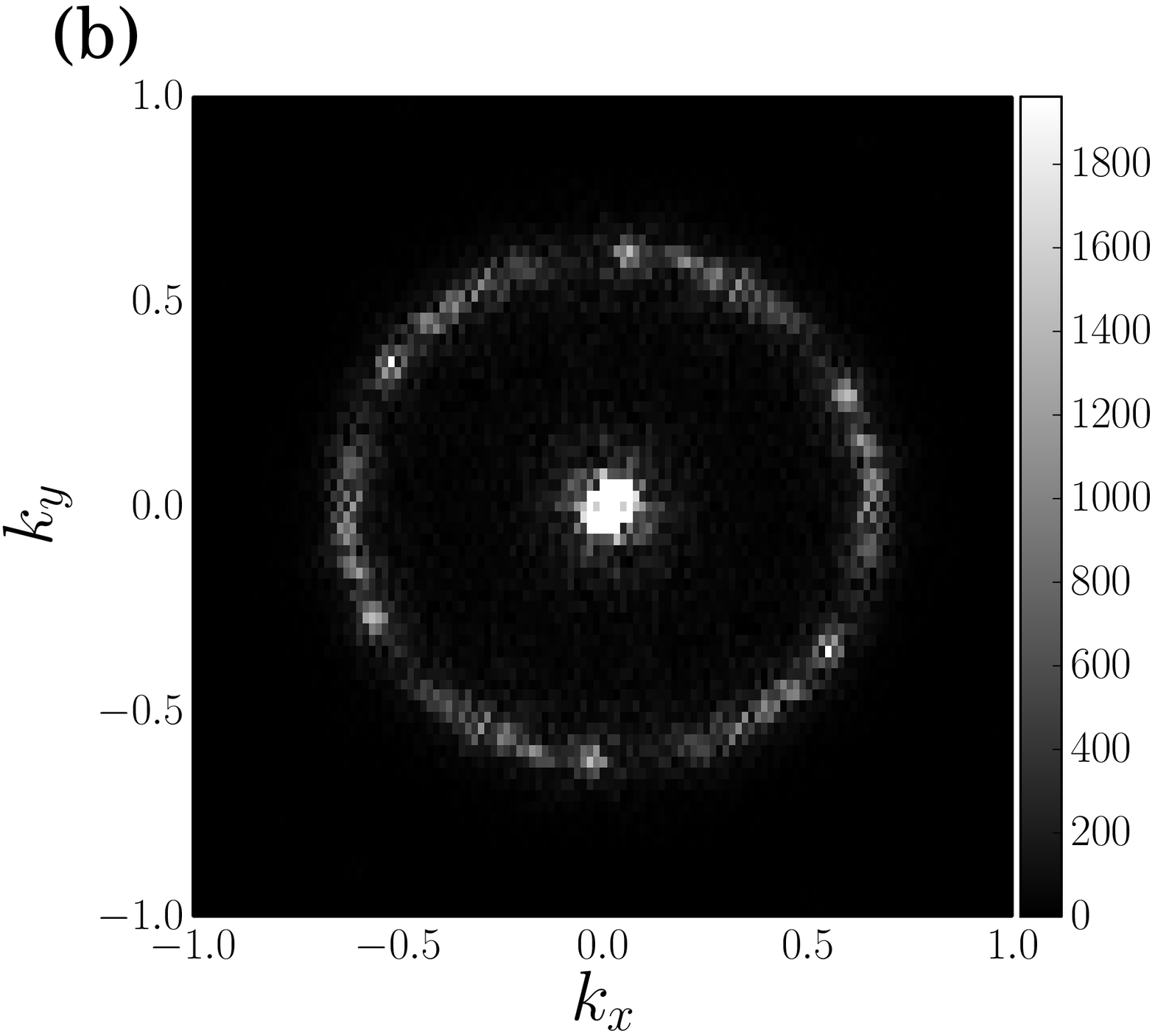}
\end{tabular}
\caption{The height (a) and the magnitude of the Fourier transform (b) of a non-templated surface after integrating to time $t=10^4$. }
\label{fig:no_templating_dots_rect}
\end{figure}

First, we present the control case in which the initial condition of the simulations was small amplitude spatial white noise; i.e., there was no templating. In Fig.~\ref{fig:no_templating_dots} (a), the surface height at time $t=10^4$ is shown; it is evident that multiple domains of hexagonally ordered nanodots have formed. The magnitude of the Fourier transform of the surface height is plotted in Fig.~\ref{fig:no_templating_dots} (b). Fig.~\ref{fig:no_templating_dots} will be compared with the simulation results in Sections~\ref{sec:sin_results} and \ref{sec:scratch_results}, since those simulations were performed on square domains. Fig.~\ref{fig:no_templating_dots_rect} is the analogue of Fig.~\ref{fig:no_templating_dots} but with the simulation performed on the same rectangular domain that will be used in the hexagonal template simulations described in Section~\ref{sec:hex_results}. The Fourier transforms in both Figs.~\ref{fig:no_templating_dots} and \ref{fig:no_templating_dots_rect} exhibit a narrow band of unstable wave vectors as a diffuse annulus with mean radius $2\pi/\lambda_T\simeq 0.61$. Although there is some structure within the two annuli, it is not very pronounced. This indicates that the hexagonal ordering is strong only locally; globally, there is no preferred orientation for the hexagons.

\subsection{Hexagonal templates}

\begin{figure}[htp]
\centering
\begin{tabular}{@{}cc@{}}
\includegraphics[width=0.45\textwidth]{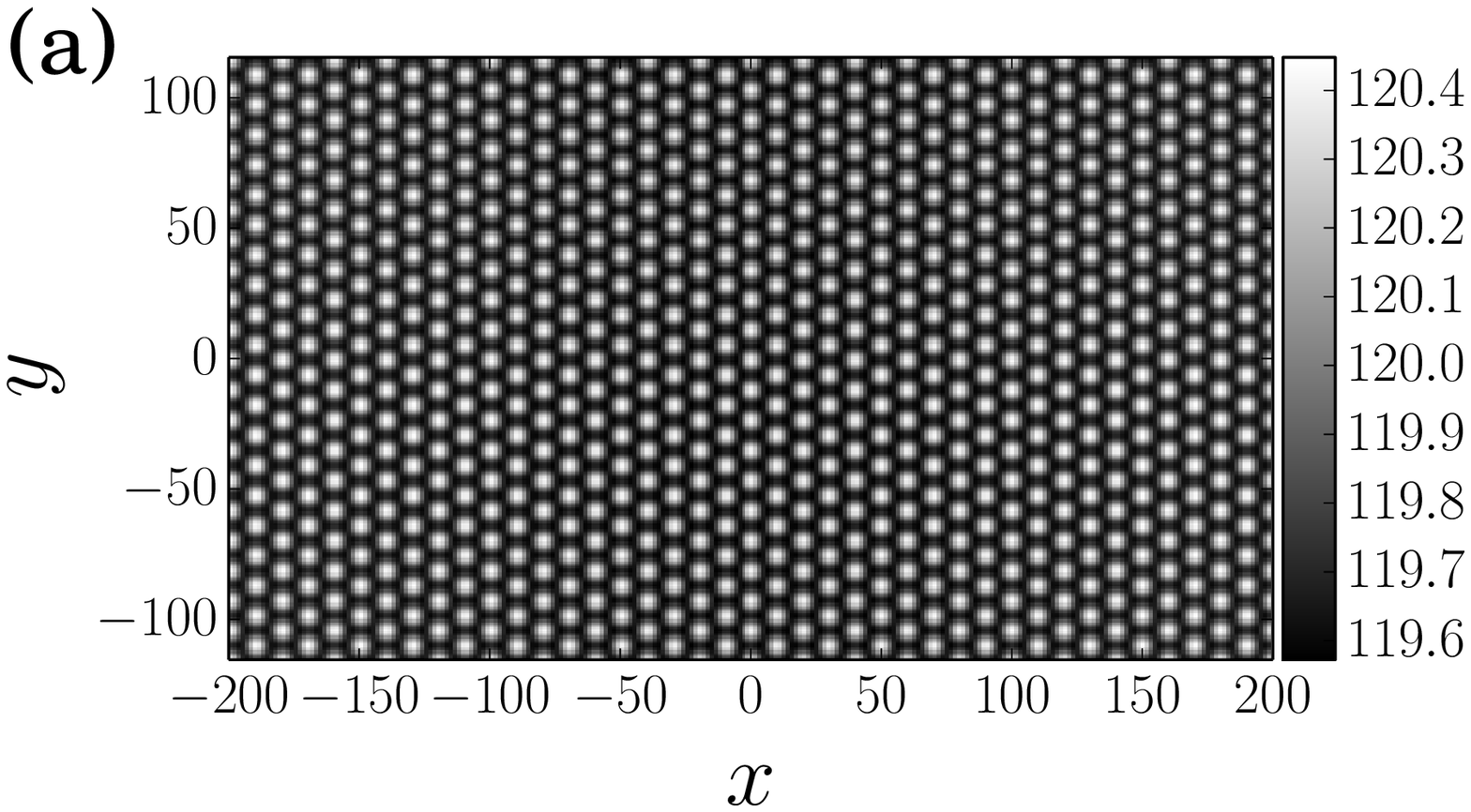} &
\includegraphics[width=0.45\textwidth]{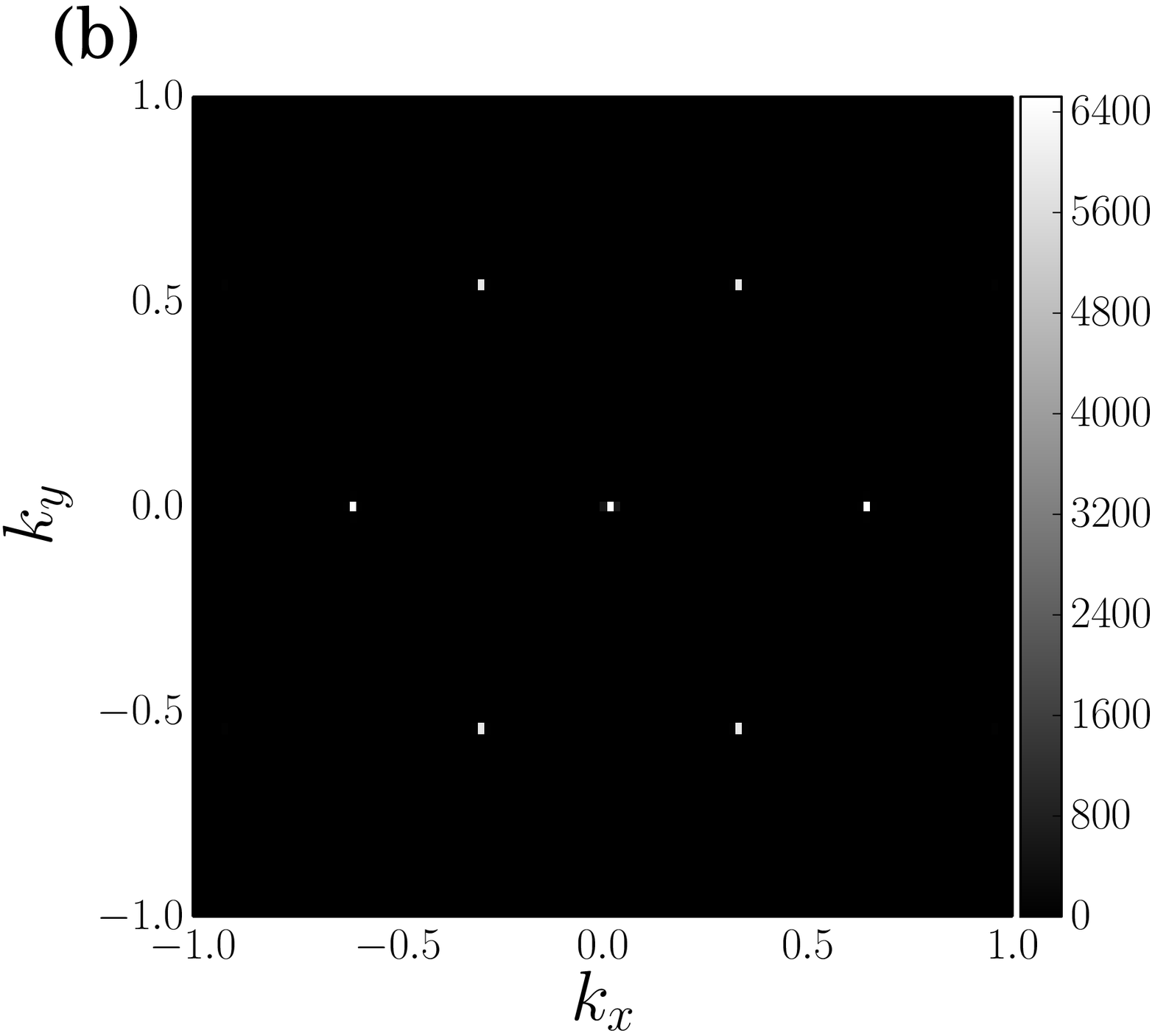}
\end{tabular}
\caption{The height (a) and the magnitude of the Fourier transform (b) of a hexagonally templated surface after integrating to time $t=10^4$. The initial wavelength was $\lambda_1=20 \simeq 2\lambda_T$. }
\label{fig:10_hex}
\end{figure}

\begin{figure}[htp]
\centering
\begin{tabular}{@{}cc@{}}
\includegraphics[width=0.45\textwidth]{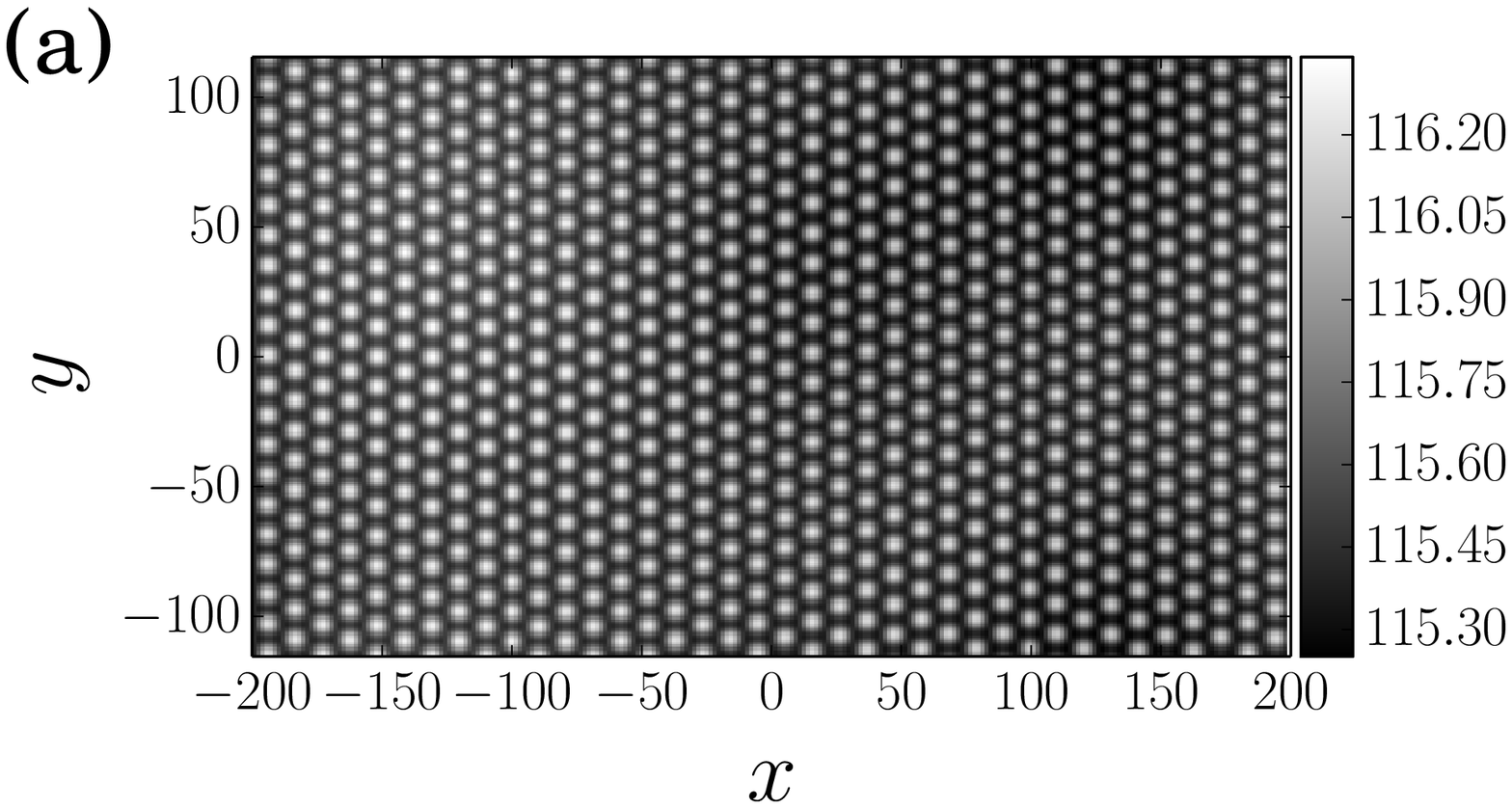} &
\includegraphics[width=0.45\textwidth]{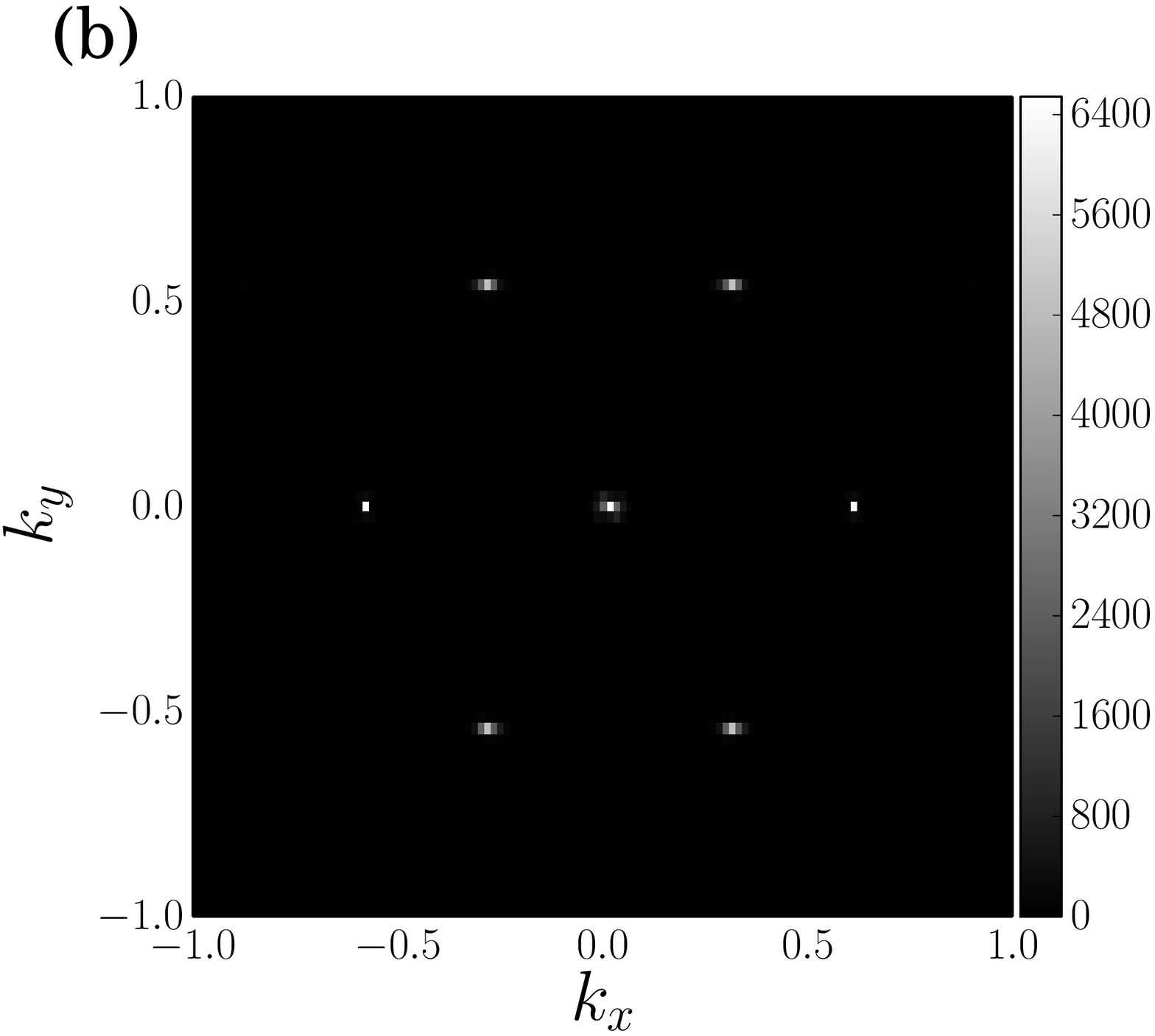}
\end{tabular}
\caption{The height (a) and the magnitude of the Fourier transform (b) of a hexagonally templated surface after integrating to time $t=10^4$. The initial wavelength was $\lambda_1=400/38 \simeq \lambda_T$.}
\label{fig:19_hex}
\end{figure}

\begin{figure}[htp]
\centering
\begin{tabular}{@{}cc@{}}
\includegraphics[width=0.45\textwidth]{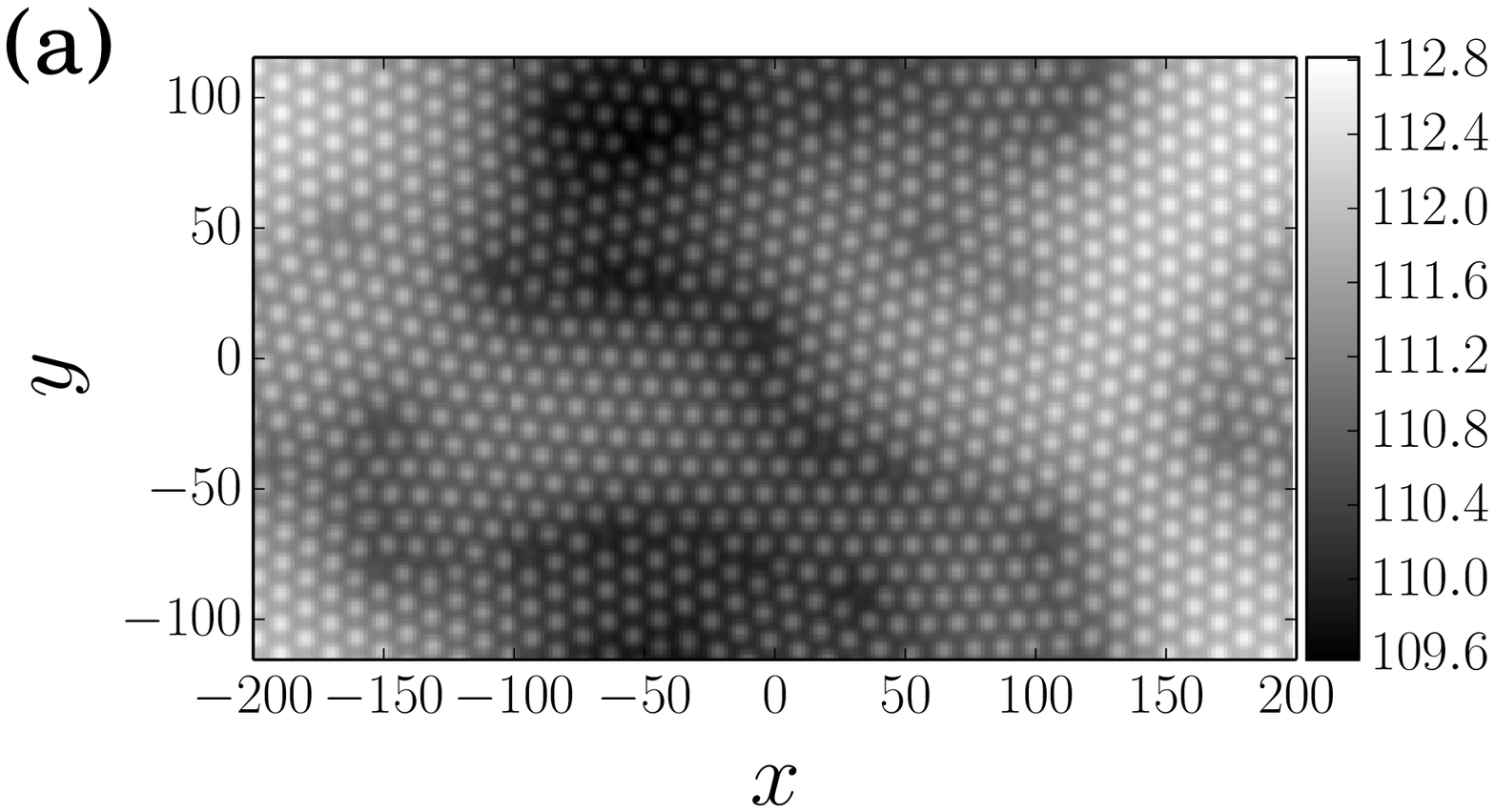} &
\includegraphics[width=0.45\textwidth]{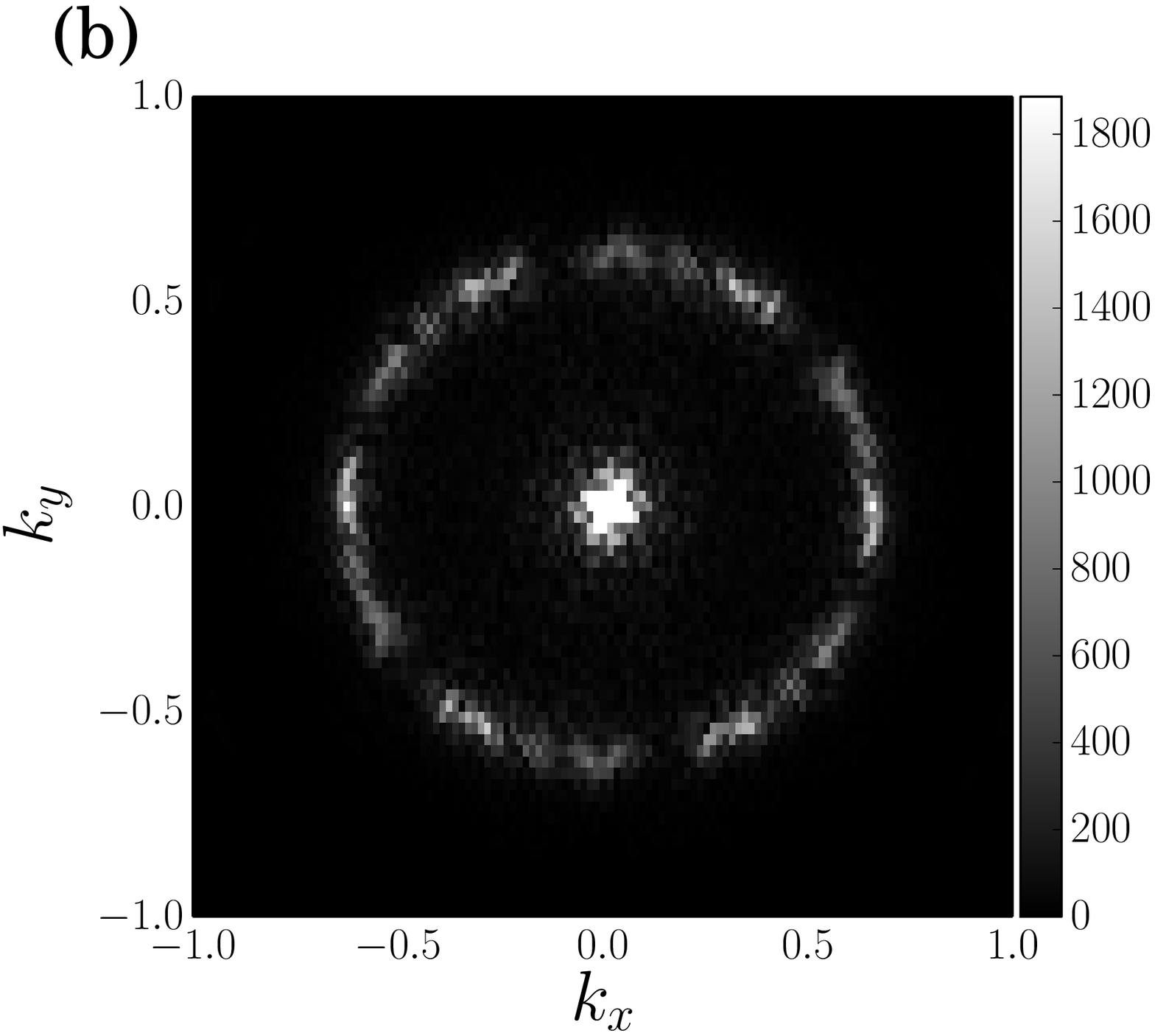}
\end{tabular}
\caption{The height (a) and the magnitude of the Fourier transform (b) of a hexagonally templated surface after integrating to time $t=10^4$. The initial wavelength was $\lambda_1=400/18\simeq 22.2$.}
\label{fig:9_hex}
\end{figure}

For the simulations using a hexagonal initial condition, we find that there can be either little effect or a dramatic improvement of the global hexagonal order of the nanodots. Dramatic improvement of the hexagonal order was only observed when the wavelength of the initial sine waves was approximately equal to $\lambda_T$ or $2\lambda_T$. For example, this effect can be seen clearly in Figs.~\ref{fig:10_hex} and \ref{fig:19_hex}, which were generated from the simulations with $\lambda_1\simeq 2\lambda_T$ and $\lambda_1\simeq\lambda_T$, respectively. The Fourier transforms of these surface heights also demonstrate the strong global hexagonal order by exhibiting six equally spaced peaks in the annulus of unstable wave vectors.  On the other hand, if $\lambda_1$ was not close to $2\lambda_T$ or $\lambda_T$, no strong improvement in order was observed, as in Fig.~\ref{fig:9_hex}. 

\begin{figure}[htp]
\centering
\includegraphics[width=0.45\textwidth]{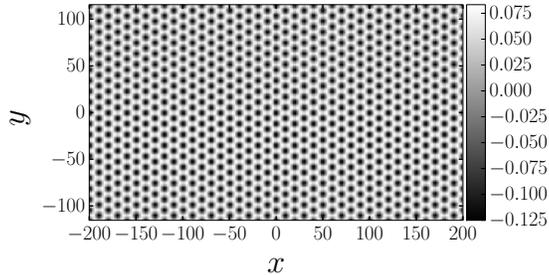}
\caption{The deviation of the surface composition from its steady-state value for the hexagonally templated surface after integrating to time $t=10^4$. The initial wavelength was $\lambda_1=400/38 \simeq \lambda_T$. The simulation is the same as the one which produced the surface height seen in Fig.~\ref{fig:10_hex} (a).}
\label{fig:10_hex_comp}
\end{figure}

Figure~\ref{fig:10_hex_comp} shows the composition corresponding to Fig.~\ref{fig:10_hex}. The surface height and composition are anticorrelated, as the Bradley-Shipman theory predicts \cite{Shipman11}. Since this is always true at sufficiently long times (including the time we ended our simulations, $t=10^4$), we will not show any additional plots of the surface composition. 

A comparison of Figs.~\ref{fig:no_templating_dots_rect} and \ref{fig:10_hex} shows that when the pattern has enhanced order, long wavelength variations in the surface height are suppressed. This effect has been investigated in detail by Motta \textit{et al.}~in the more general case of obliquely-incident ion bombardment of binary materials \cite{oblique}. Using the amplitude equations they derived, Motta \textit{et al.}~found that average height of a region evolves differently depending on whether the region has defects. Specifically, a region with defects will typically be eroded faster than a defect-free region, which is in agreement with our simulation results.
\label{sec:hex_results}

\begin{figure}[htp]
\centering
\includegraphics[width=0.5\textwidth]{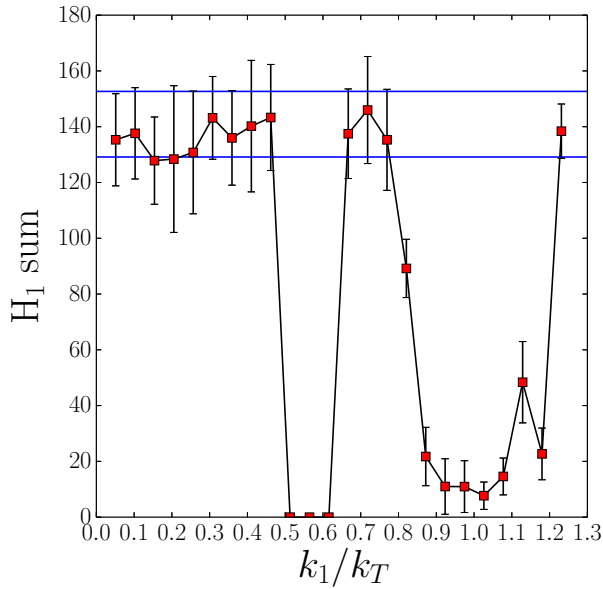}
\caption{(color online) The $H_1$ sum versus the ratio $k_1/k_T$ for the hexagonal templates after integrating to time $t=10^4$, averaged over 10 realizations. The two horizontal blue lines show the $H_1$ sum averaged over 10 non-templated initial surfaces after integrating to time $t=10^4$, plus or minus the standard deviation.}
\label{fig:H1_sum_hex}
\end{figure}

In order to quantify the dependence of global hexagonal order on the initial wavelength $\lambda_1$, we simulated the evolution of the surface for 24 different initial wavelengths. Furthermore, we performed 10 simulations at each of these wavelengths. Using the persistent homology method of Section~\ref{ph_order}, we computed $H_1$ sums for each of the initial wavelengths, and then averaged the results over the 10 realizations. In all our $H_1$ sum calculations, we filtered out persistence intervals with lengths less than the one pixel resolution of the local maximum finder.  The results are shown in Fig.~\ref{fig:H1_sum_hex}. The error bars were obtained from the standard deviations of the 10 trials at each wavelength. The $H_1$ sum characterizes the hexagonal order in a way that agrees with how one would qualitatively describe the order based on a visual inspection of the real space results. In particular, it shows the excellent global hexagonal order that occurs for $\lambda_1\simeq 2\lambda_T$ and $\lambda_1\simeq\lambda_T$. There is also a much larger $H_1$ sum for the simulations which had $\lambda_1\simeq 22.2$, or equivalently $k_1/k_T\simeq 0.46$. One such simulation is shown in Fig.~\ref{fig:9_hex}. A visual inspection corroborates the $H_1$ sum's indication that the hexagonal ordering of these surface is on par with the results of the non-templated surfaces, such as Fig.~\ref{fig:no_templating_dots_rect}. Furthermore, there is improved order for many simulations with initial wavelengths near the linearly selected wavelength. This is also observed in the real space results, such as Fig.~\ref{fig:19_hex}. \\

\begin{figure}[htp]
\centering
\includegraphics[width=0.5\textwidth]{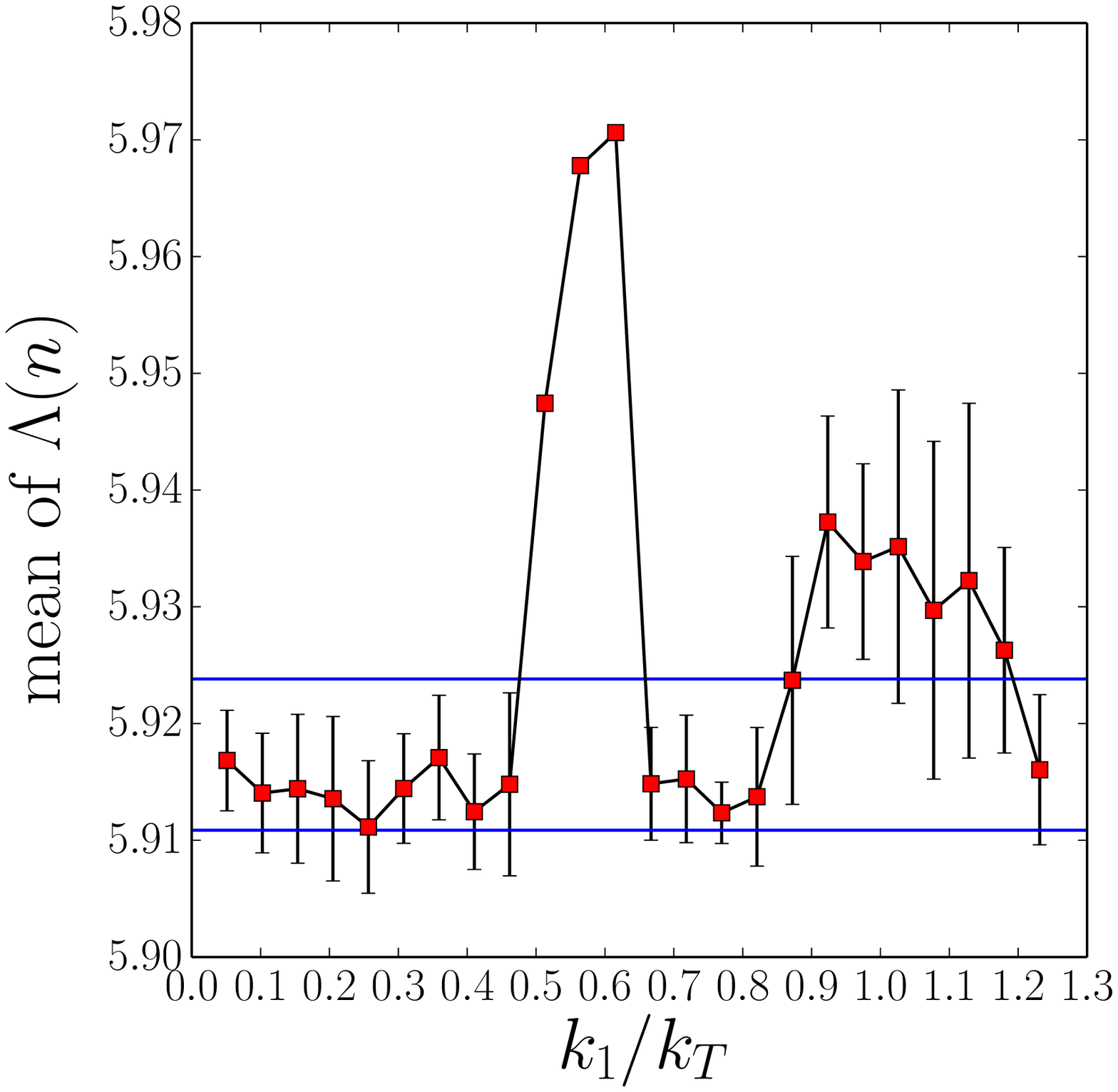}
\caption{(color online) The mean of the nearest-neighbor distribution $\Lambda(n)$ versus the ratio $k_1/k_T$ for the hexagonal templates after integrating to time $t=10^4$, averaged over 10 realizations. The two horizontal blue lines show the mean of $\Lambda(n)$ averaged over 10 non-templated initial surfaces after integrating to time $t=10^4$, plus or minus the standard deviation.}
\label{fig:voronoi_avg_hex}
\end{figure}

\begin{figure}[htp]
\centering
\includegraphics[width=0.5\textwidth]{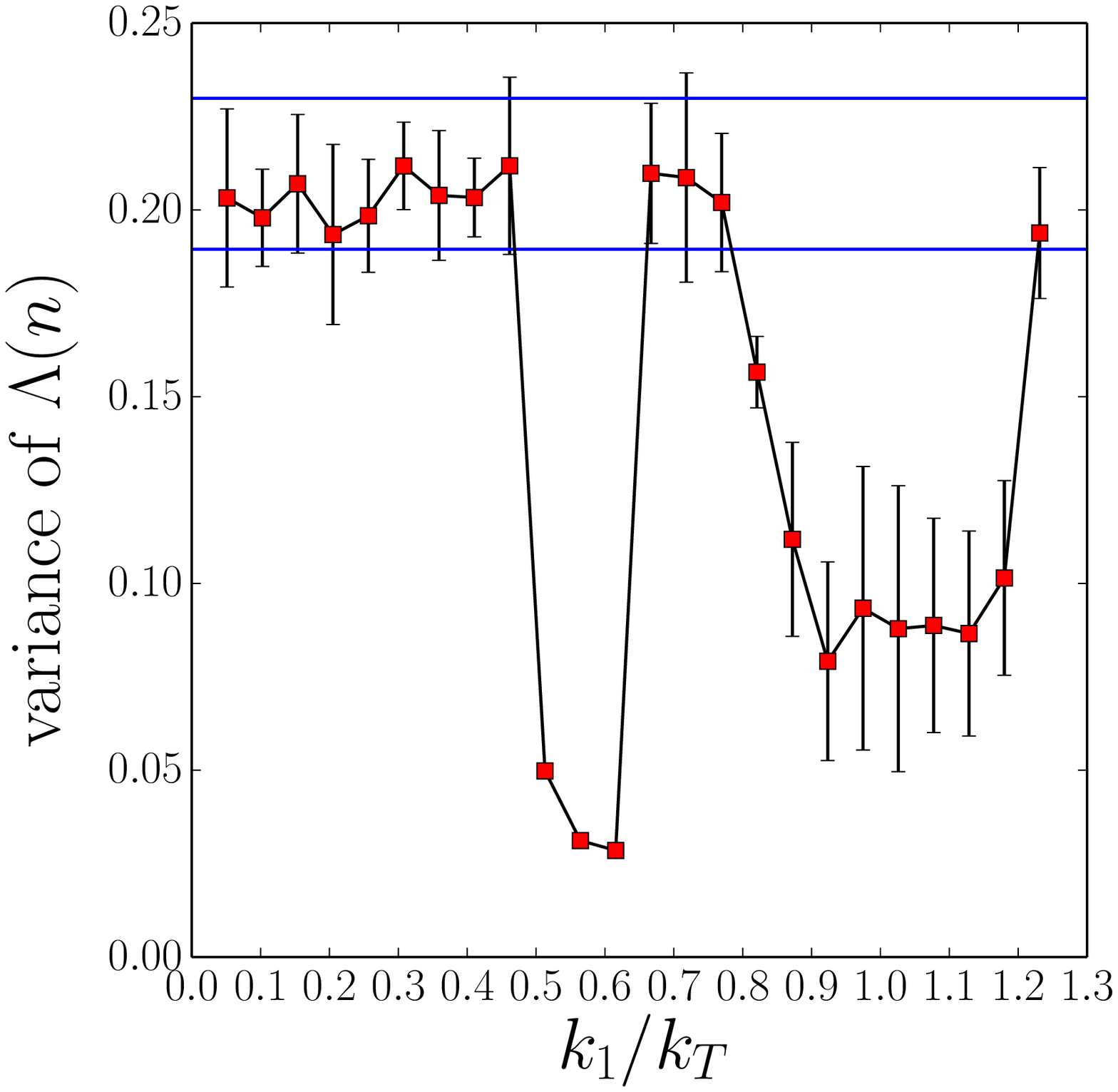}
\caption{(color online) The variance of the nearest-neighbor distribution $\Lambda(n)$ versus the ratio $k_1/k_T$ for the hexagonal templates after integrating to time $t=10^4$, averaged over 10 realizations. The two horizontal blue lines show the variance of $\Lambda(n)$ averaged over 10 non-templated initial surfaces after integrating to time $t=10^4$, plus or minus the standard deviation.}
\label{fig:voronoi_var_hex}
\end{figure}

We further analyzed the hexagonal templates using the nearest-neighbor number distribution introduced in Section~\ref{nn_order}. For the results, see Figs.~\ref{fig:voronoi_avg_hex} and \ref{fig:voronoi_var_hex}. Recall that for a perfectly hexagonal lattice the mean and variance of $\Lambda(n)$ would be 6 and 0, respectively. The results are qualitatively in agreement with those obtained using the $H_1$ sum. The advantage of the $H_1$ sum over the Voronoi method is that the $H_1$ sum plot clearly shows that each of the simulations with $\lambda_1\simeq 2\lambda_T$ evolved to a perfectly ordered hexagonal array of nanodots, while the Voronoi plots only indicate improved order for those simulations. This occurs despite the fact that the $H_1$ sum is more sensitive to small perturbations than $\Lambda(n)$ \cite{ph_motta_shipman}. The reason the $H_1$ sum identifies the perfectly ordered hexagonal arrays and the Voronoi method does not is that we could filter out noise caused by the finite resolution of the local maximum finder when calculating the $H_1$ sum but not when calculating $\Lambda(n)$.

Do templates of greater amplitude lead to patterns with a lower degree of order?  To address the this question, we again carried out simulations with $\lambda_1=400/38 \simeq \lambda_T$ and with $\lambda_1=20 \simeq 2\lambda_T$.  The amplitude of the template, however, was increased by a factor of 10 to 0.1.  The amplitude of the low amplitude spatial white noise was left unchanged and the degree of hexagonal order which was once again measured using the $H_1$ sum.  Our simulations show that for both values of $\lambda_1$, the quality of the hexagonal order was undiminished by the ten-fold increase in the template amplitude.


\subsection{Sinusoidal templates}
\label{sec:sin_results}
\begin{figure}[htp]
\centering
\begin{tabular}{@{}cc@{}}
\includegraphics[width=0.45\textwidth]{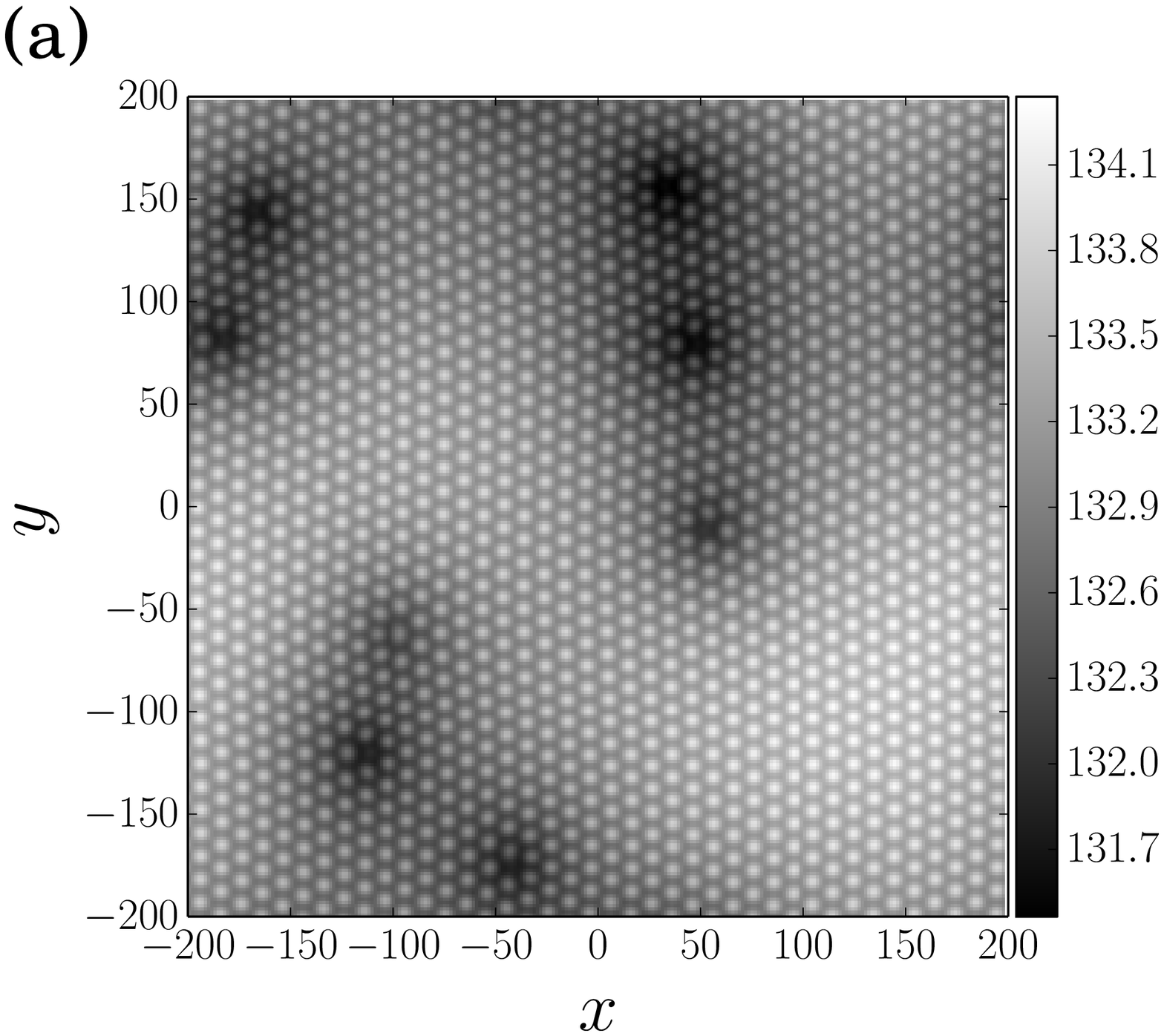} &
\includegraphics[width=0.45\textwidth]{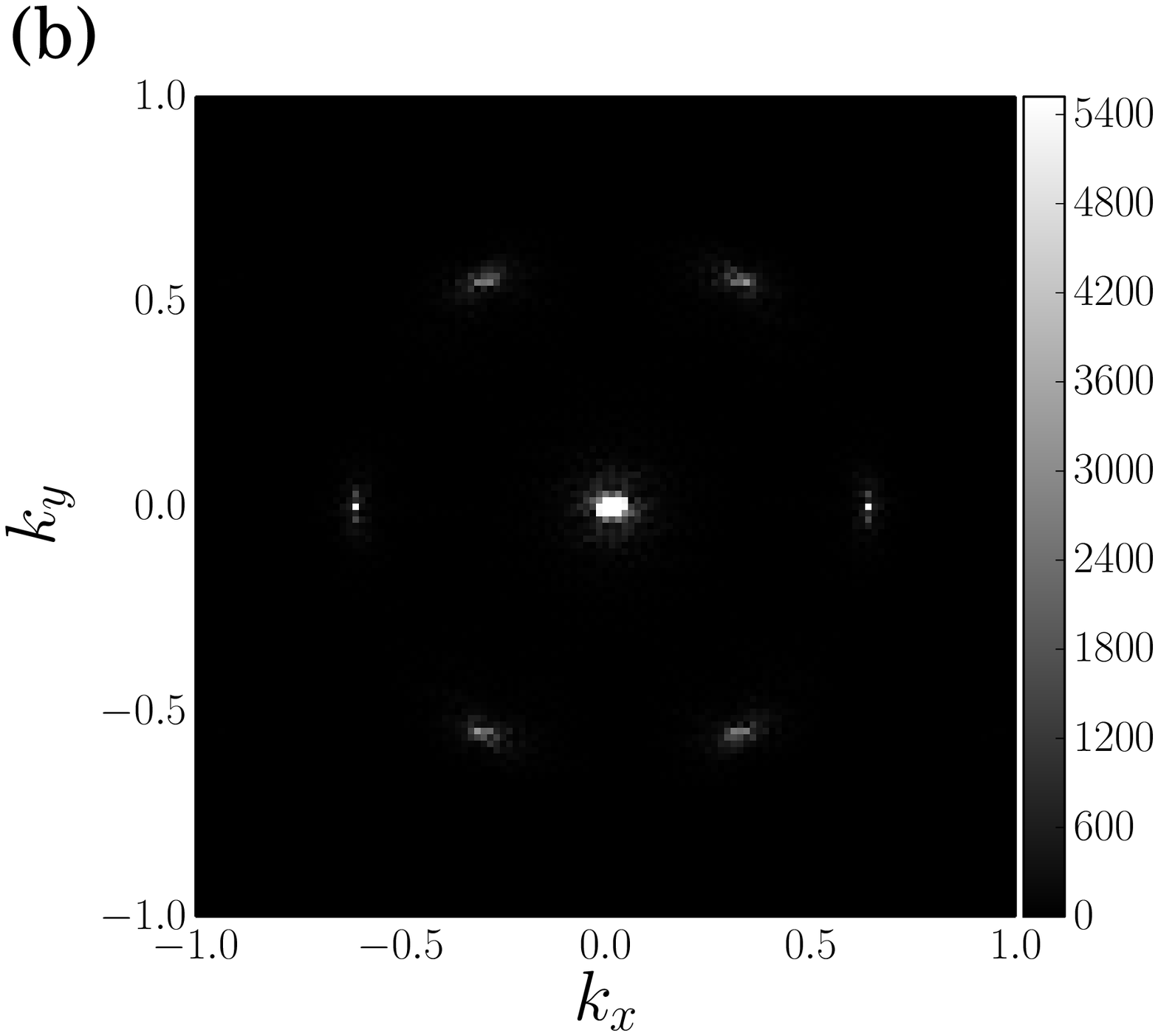}
\end{tabular}
\caption{The surface height (a) and the magnitude of the Fourier transform (b) after integrating to time $t=10^4$ for a sinusoidal template with $\lambda_2=20\simeq 2\lambda_T$.}
\label{fig:20_dots}
\end{figure}

\begin{figure}[htp]
\centering
\begin{tabular}{@{}cc@{}}
\includegraphics[width=0.45\textwidth]{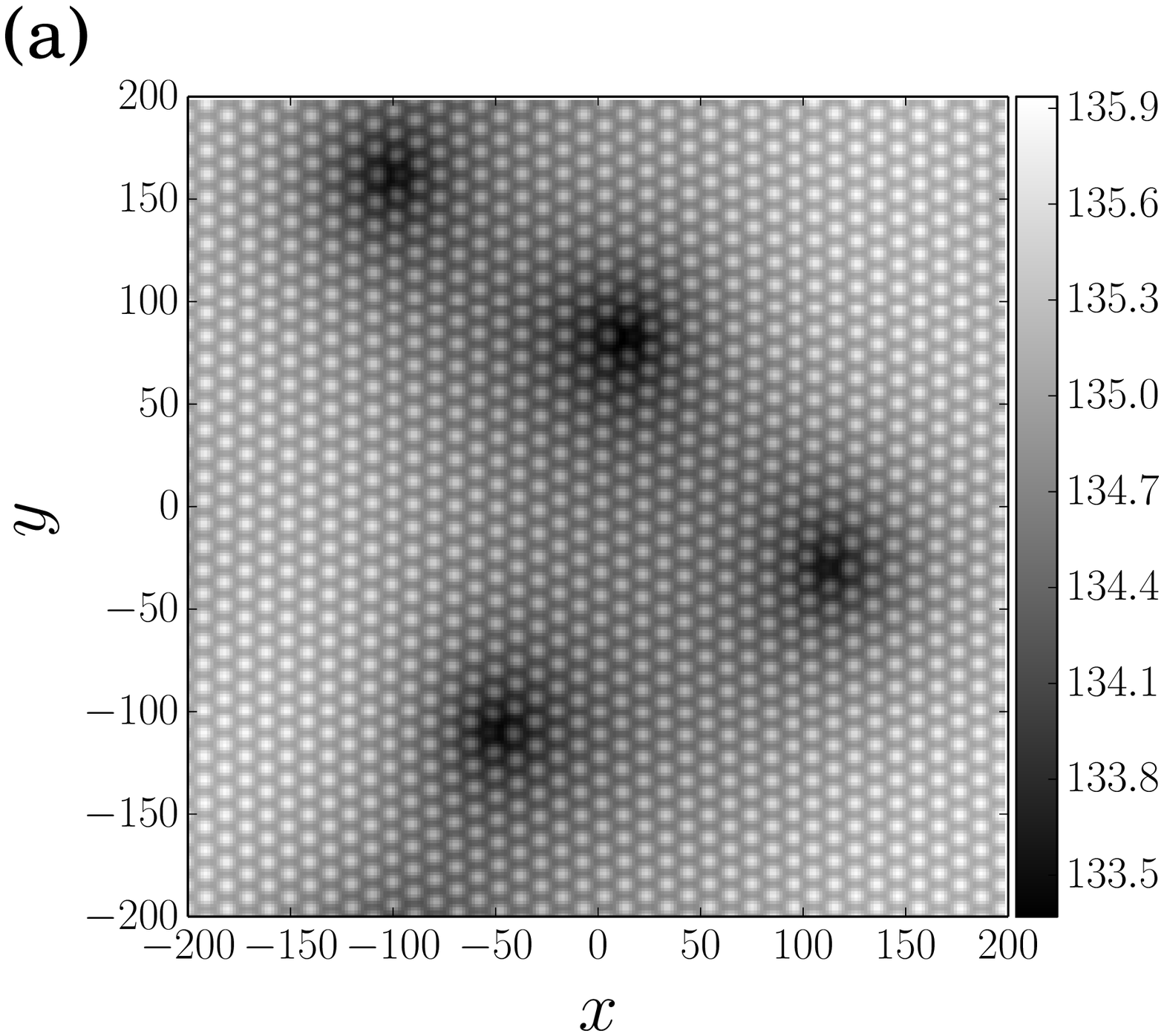} &
\includegraphics[width=0.45\textwidth]{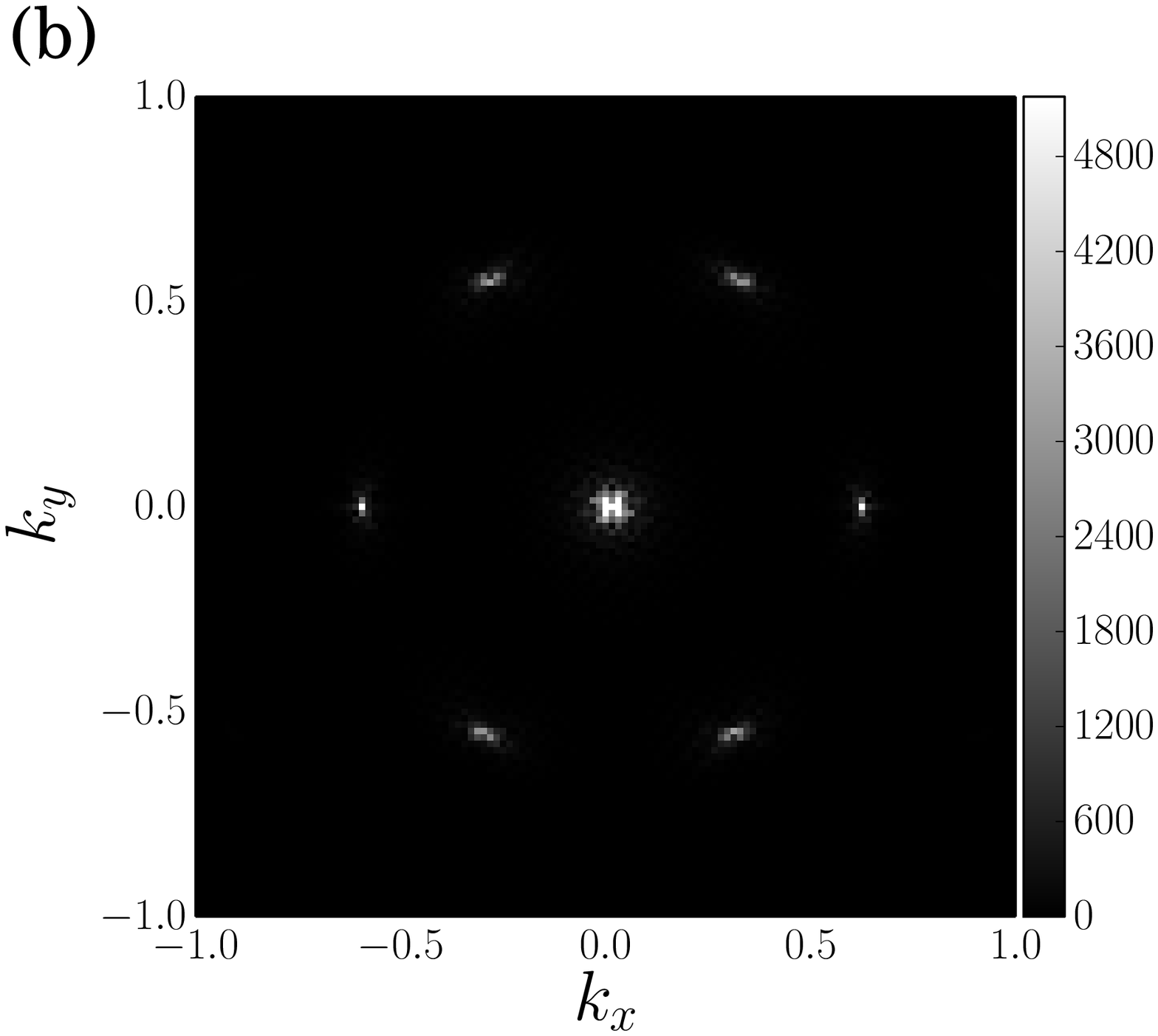}
\end{tabular}
\caption{The surface height (a) and the magnitude of the Fourier transform (b) after integrating to time $t=10^4$ for a sinusoidal template with $\lambda_2=400/39\simeq \lambda_T$. }
\label{fig:39_dots}
\end{figure}

For the simulations that began with a sinusoidal initial condition, again dramatic improvement of the hexagonal order was only seen when the wavelength of the initial sine wave was approximately equal to $\lambda_T$ or $2\lambda_T$. For example, this effect can be seen clearly in Figs.~\ref{fig:20_dots} and \ref{fig:39_dots}, which were generated from the simulations with $\lambda_2\simeq 2\lambda_T$ and $\lambda_2\simeq\lambda_T$, respectively. The Fourier transforms of these surface heights also demonstrate the strong global hexagonal order by exhibiting six equally spaced peaks in the annulus of unstable wave vectors. On the other hand, if $\lambda_2$ was not close to $2\lambda_T$ or $\lambda_T$, no strong improvement in order was observed.


\subsection{Scratched templates}
\label{sec:scratch_results}
\begin{figure}[htp]
\centering
\begin{tabular}{@{}cc@{}}
\includegraphics[width=0.45\textwidth]{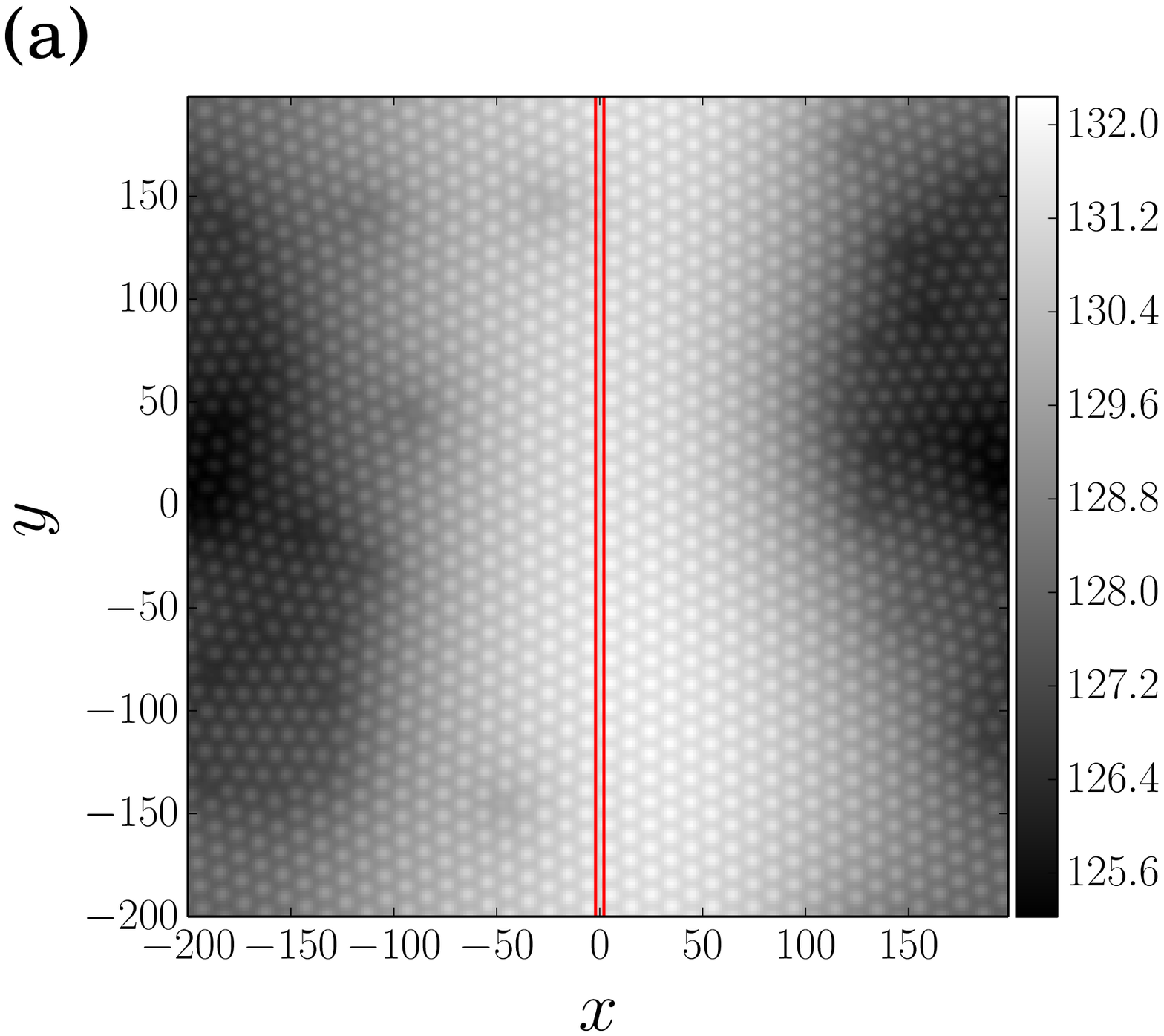} &
\includegraphics[width=0.45\textwidth]{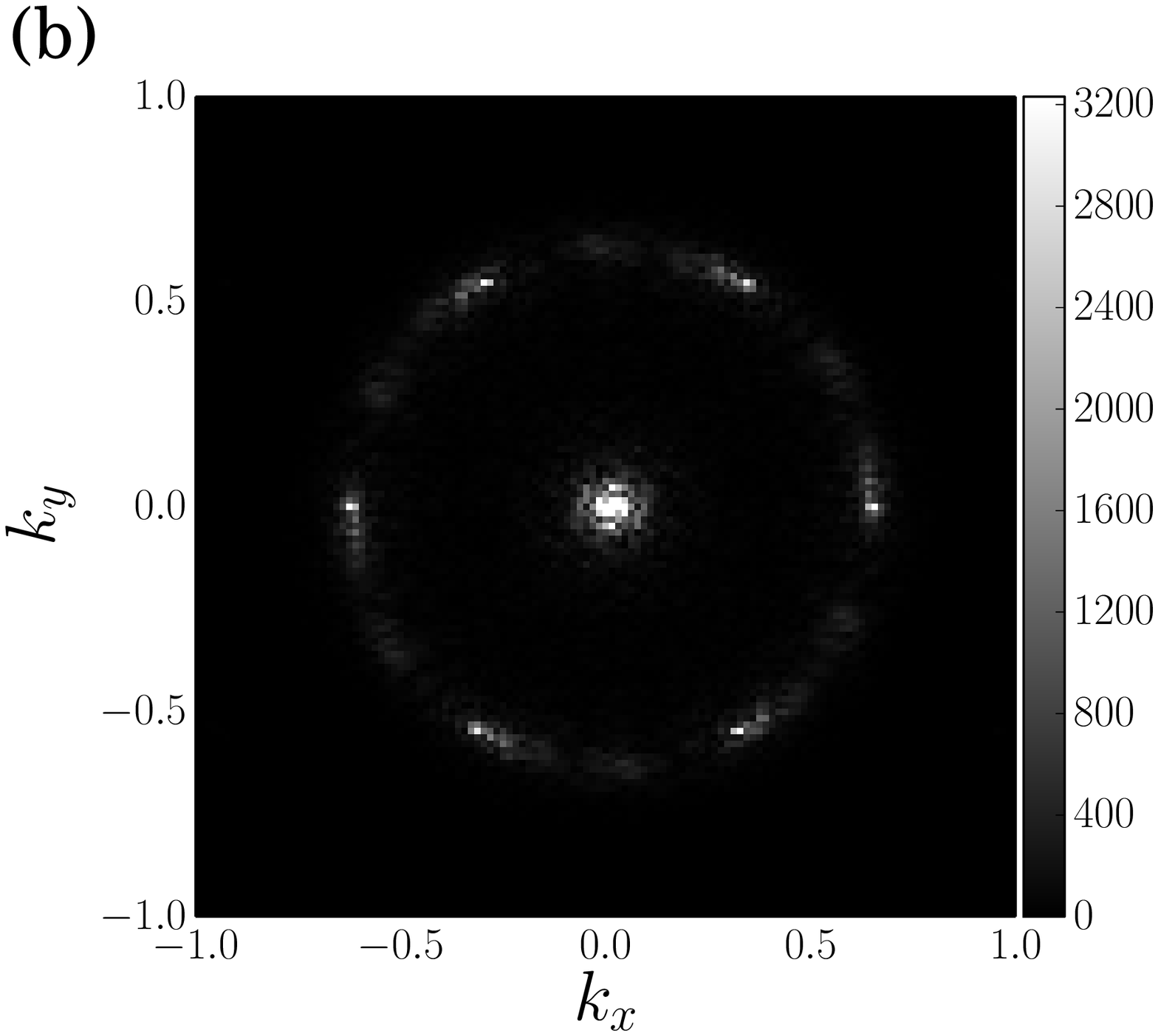}
\end{tabular}
\caption{(color online) Surface height (a) and the magnitude of the Fourier transform (b) after integrating to time $t=10^4$ with a scratch of width $2\sigma=4$. The red vertical lines indicate the approximate boundary of the initial scratch. }
\label{fig:5_scratch}
\end{figure}

\begin{figure}[htp]
\centering
\begin{tabular}{@{}cc@{}}
\includegraphics[width=0.45\textwidth]{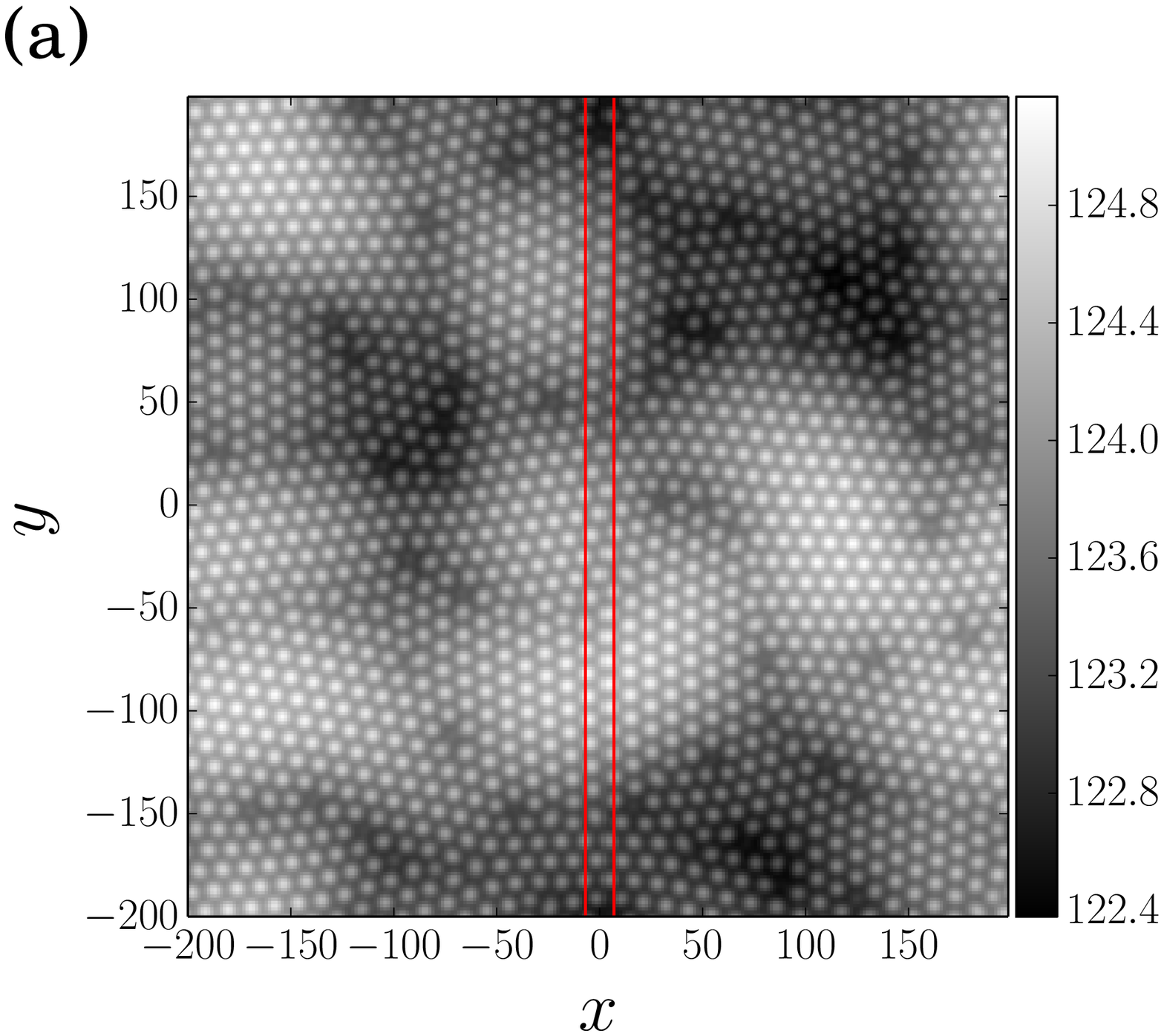} &
\includegraphics[width=0.45\textwidth]{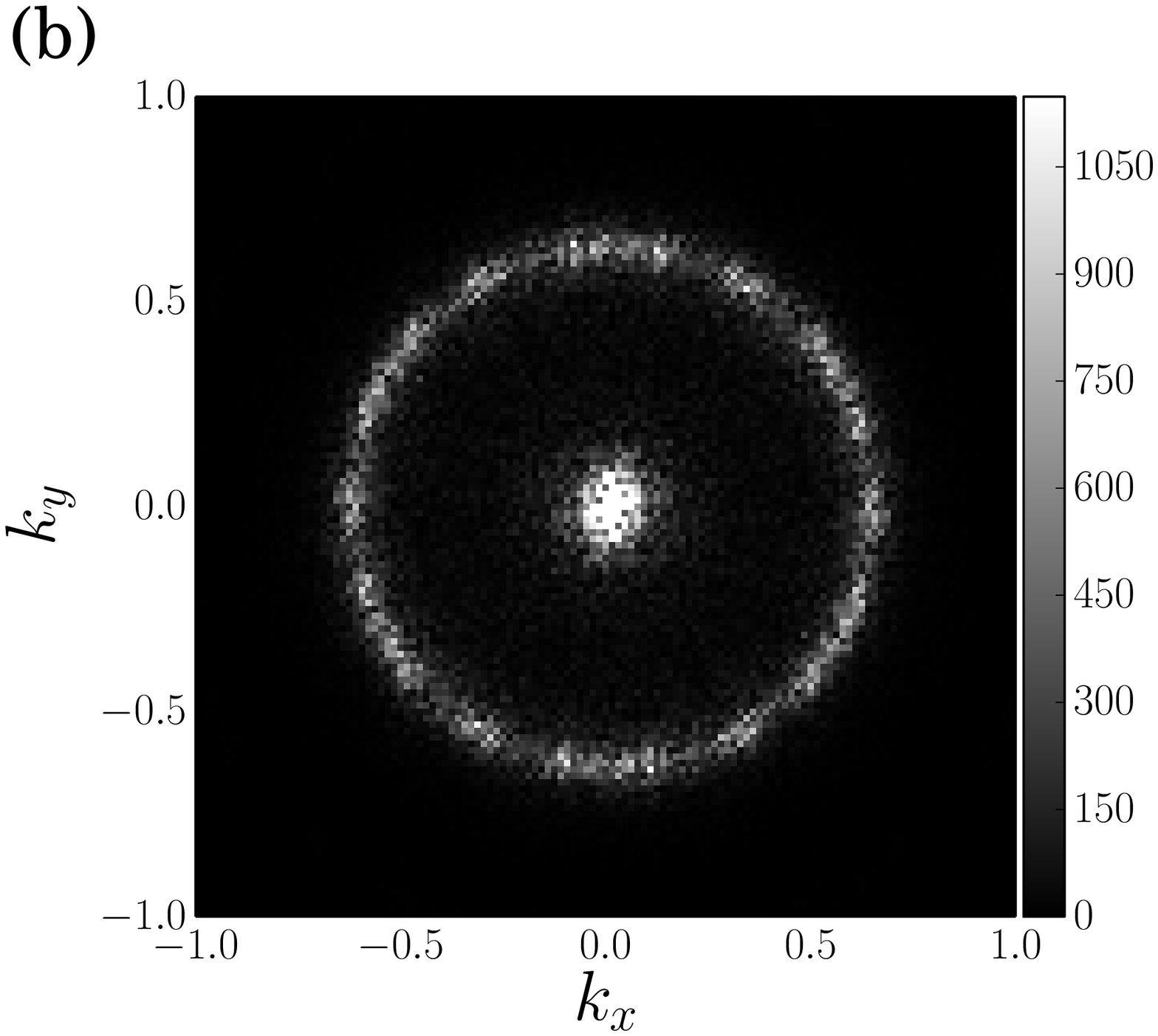}
\end{tabular}
\caption{(color online) Surface height (a) and the magnitude of the Fourier transform (b) after integrating to time $t=10^4$ with a scratch of width $2\sigma \simeq 13.86$. The red vertical lines indicate the approximate boundary of the initial scratch. }
\label{fig:60_scratch}
\end{figure}

\begin{figure}[htp]
\centering
\begin{tabular}{@{}cc@{}}
\includegraphics[width=0.45\textwidth]{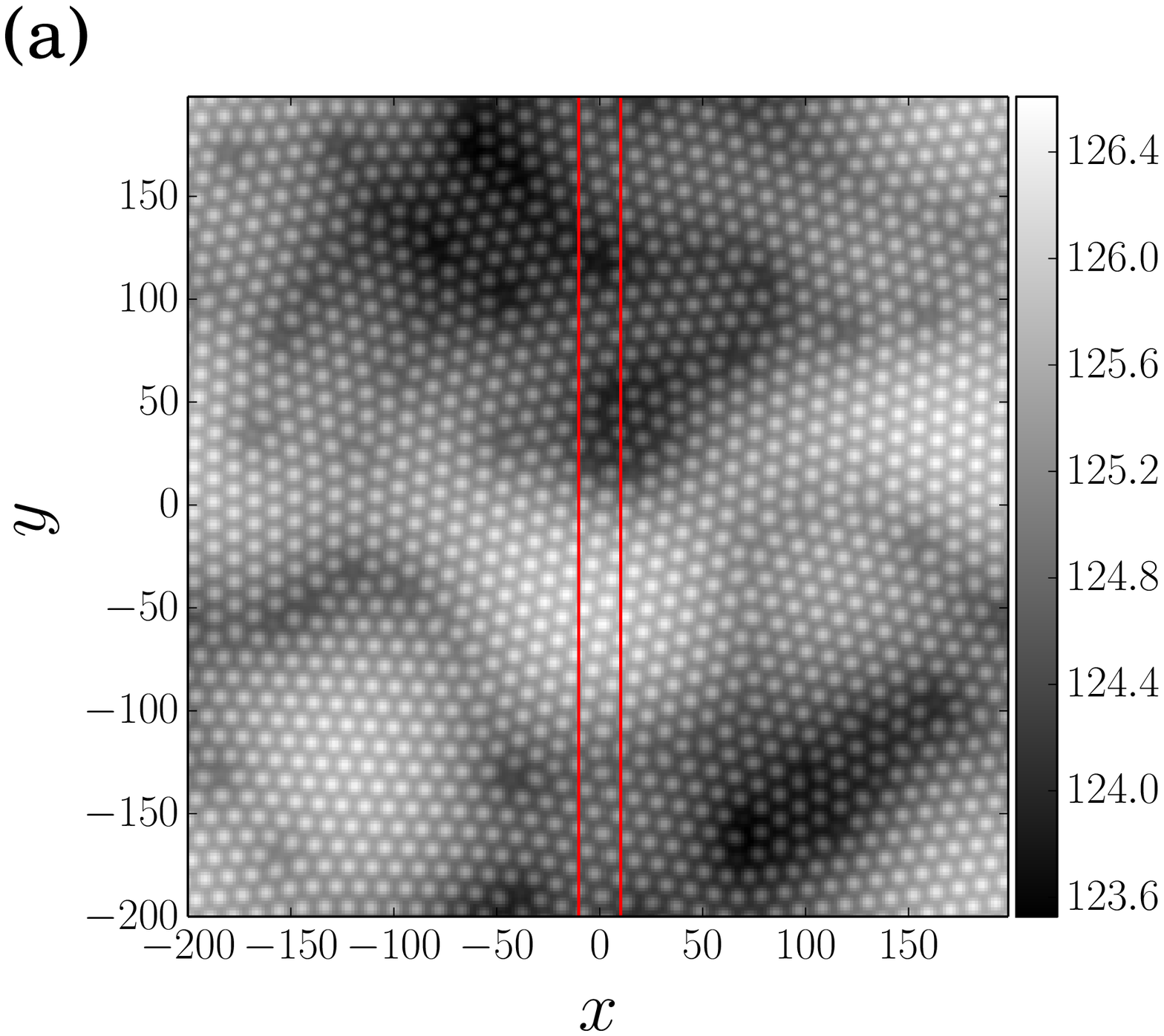} &
\includegraphics[width=0.45\textwidth]{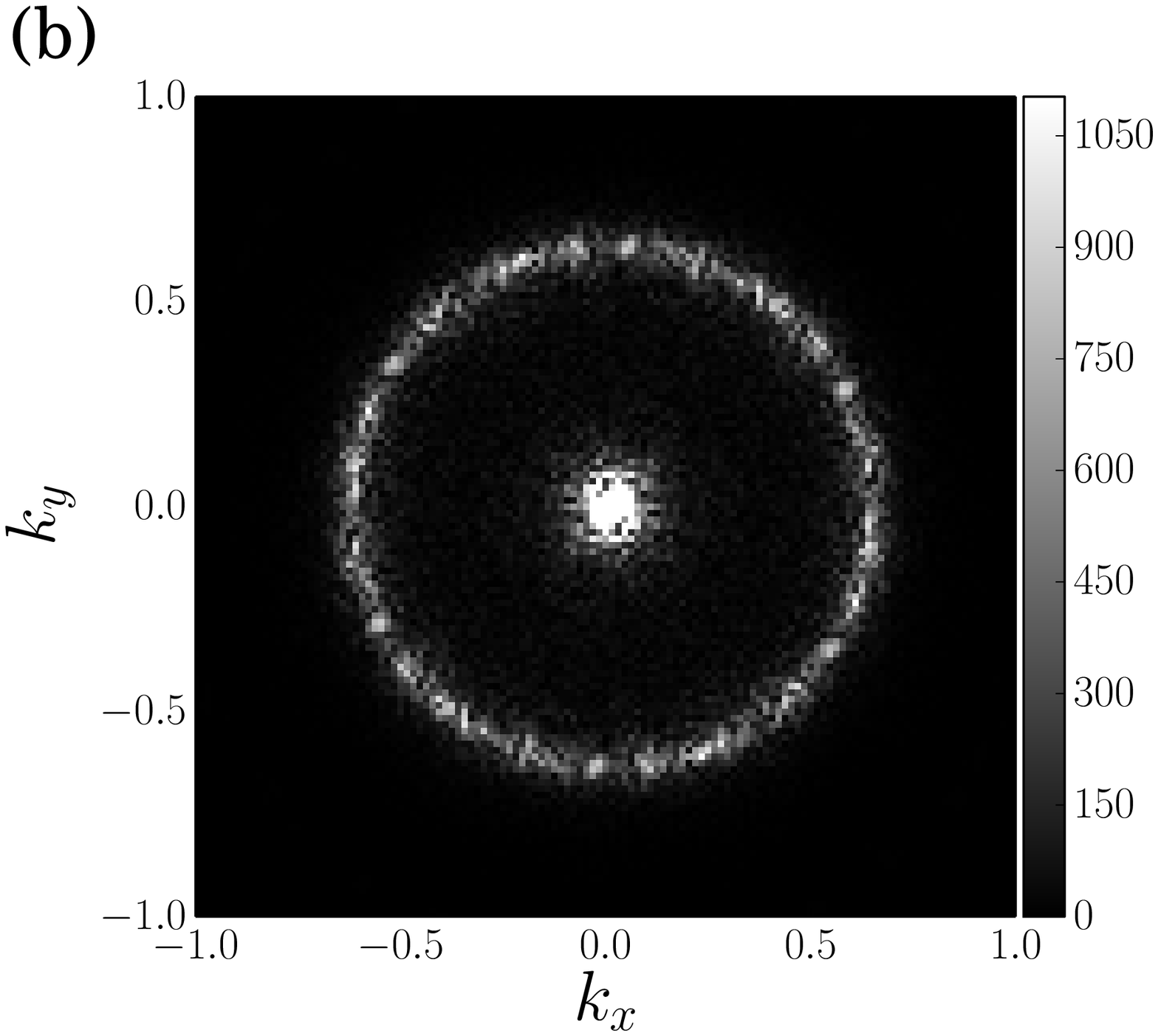}
\end{tabular}
\caption{(color online) Surface height (a) and the magnitude of the Fourier transform (b) after integrating to time $t=10^4$ with a scratch of width $2\sigma \simeq 20.40$. The red vertical lines indicate the approximate boundary of the initial scratch. }
\label{fig:130_scratch}
\end{figure}

If we start with a scratch initial condition, again improved ordering can be observed. In the case of scratch initial conditions, however, the ordering is localized along a strip centered on the initial scratch, as in Fig.~\ref{fig:5_scratch}. Furthermore, this improved ordering lasts for the full duration of the simulation ($t=10^4$) only if the width of the scratch $2\sigma$ is close to or less than the linearly selected wavelength. The real space surface produced after integrating to $t=10^4$ starting from a scratch of width 4 is shown in Fig.~\ref{fig:5_scratch} (a). The corresponding Fourier transform exhibits 6 peaks separated by $60^{\mathrm{o}}$ in the annulus of unstable wave vectors, as expected for a surface with global hexagonal order. However, if the initial scratch width is substantially larger than the linearly selected wavelength, as was the case in Fig.~\ref{fig:60_scratch} with $2\sigma\simeq 13.86$, then there is no improvement in the global hexagonal order at time $t=10^4$. The Fourier transform substantiates this claim since it exhibits a diffuse annulus devoid of any noticeable peaks. Even if the cut width is chosen to be approximately twice the linearly selected wavelength, there is still no substantial improvement in the global hexagonal order at time $t=10^4$, as is seen in Fig.~\ref{fig:130_scratch}.

The region of enhanced hexagonal order in Fig.~\ref{fig:5_scratch} (a) is higher than the remainder of the surface.  This is once again in agreement with Motta \textit{et al.}'s prediction that regions with defects are eroded at a different rate than well-ordered regions~\cite{oblique}.


\begin{figure}[htp]
\centering
\begin{tabular}{@{}cc@{}}
\includegraphics[width=0.45\textwidth]{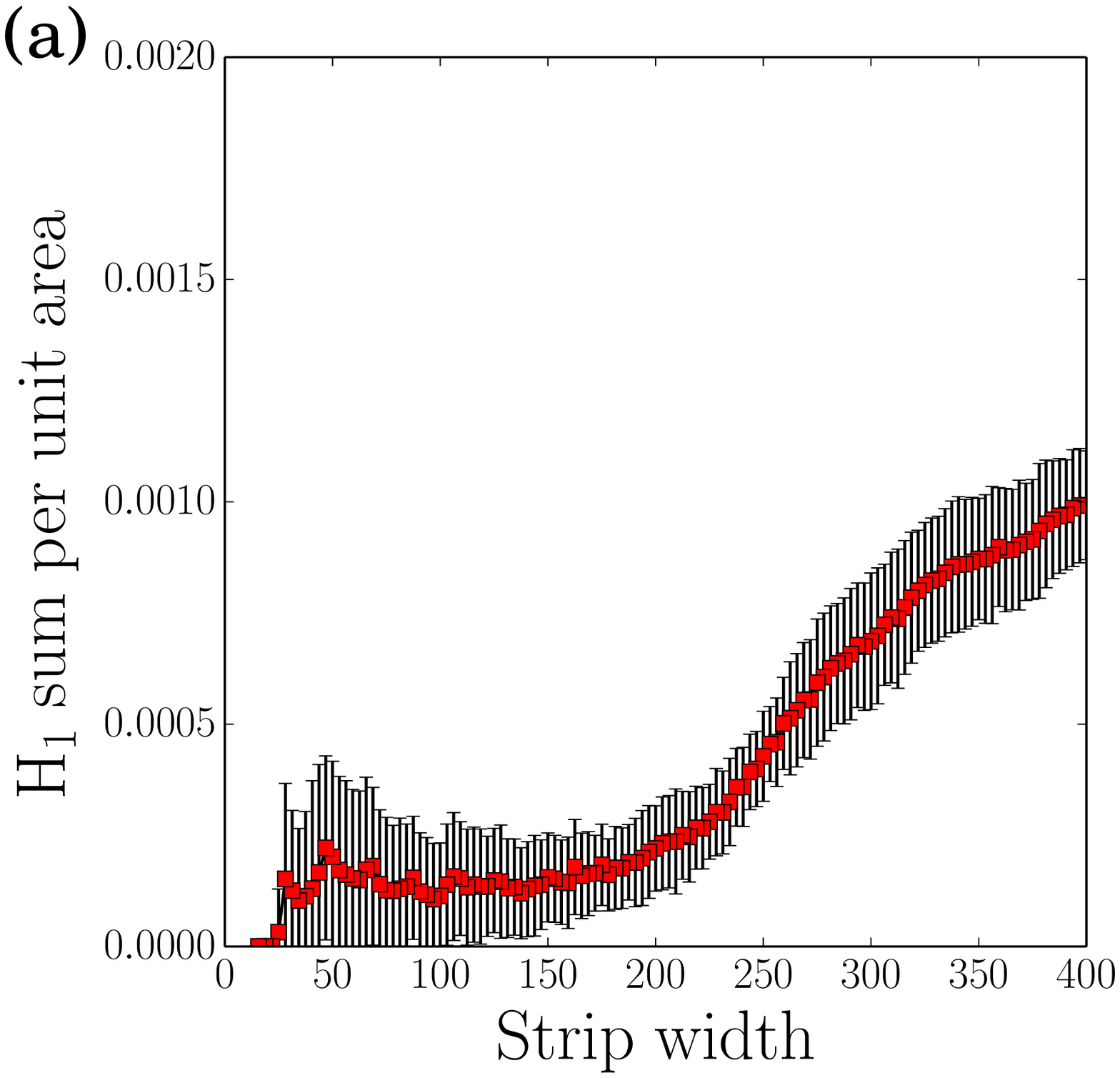} &
\includegraphics[width=0.45\textwidth]{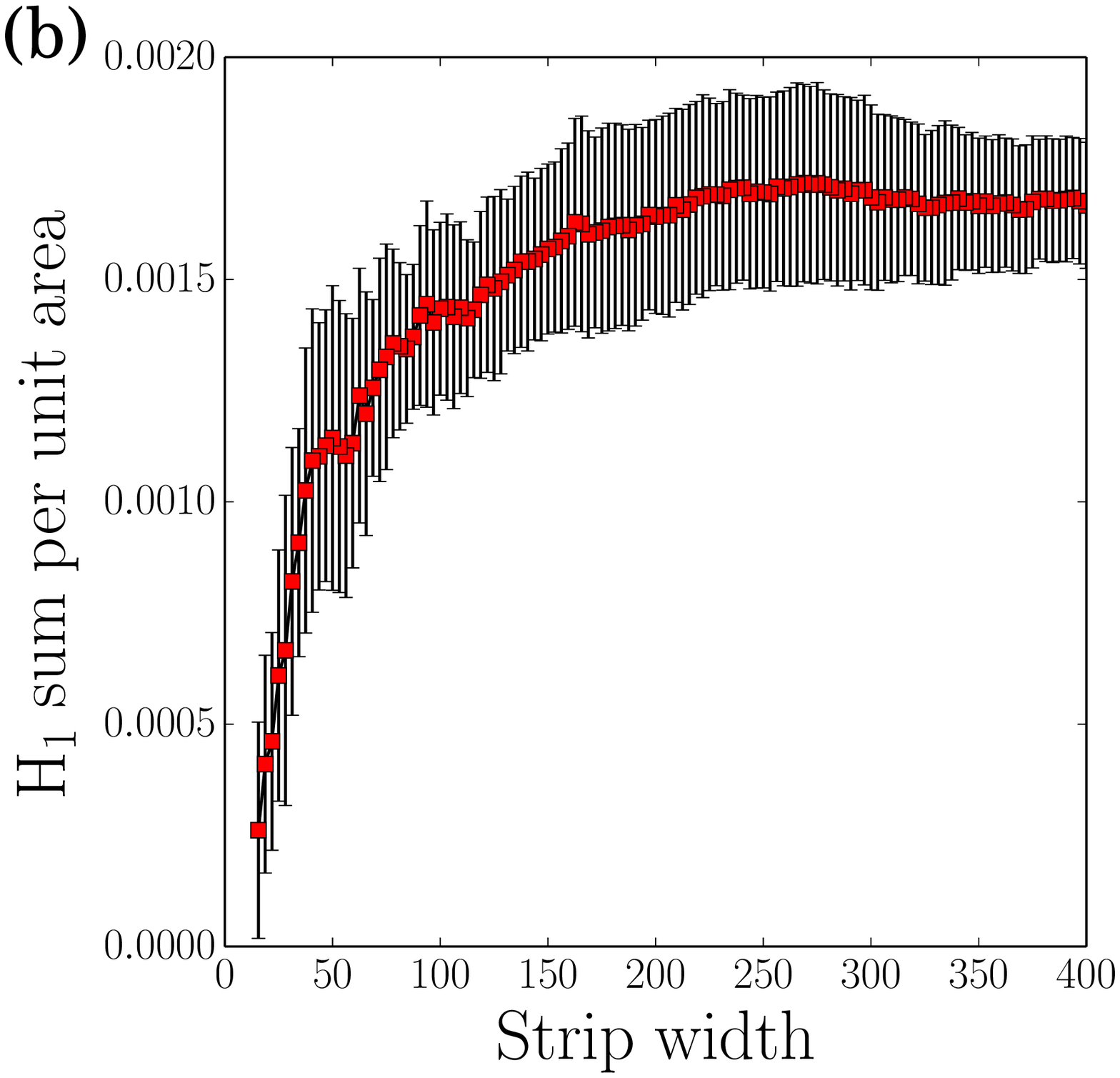}
\end{tabular}
\caption{(color online) The $H_1$ sum per unit area versus strip width for the initial conditions with a scratch of width 4 (a) and with a scratch of width approximately $2\lambda_T$ (b) after integrating to time $t=10^4$, averaged over 10 realizations.}
\label{fig:H1_sum_scratch_comb}
\end{figure}

From Fig.~\ref{fig:5_scratch} (a) it can be seen that, when the improved ordering occurs, it is most dramatic in a strip centered along the initial scratch. To measure this localization of the order quantitatively, we used the persistent homology method of Section~\ref{ph_order} on strips of different widths. Since changing the strip width also changes the total area over which one is measuring holes, it is more appropriate to calculate the $H_1$ sum per unit area instead of the raw $H_1$ sum. The result of this analysis after averaging over 10 simulations each with an initial scratch of width 4 is shown in Fig.~\ref{fig:H1_sum_scratch_comb} (a). The result from averaging 10 simulations with scratch width approximately equal to $2\lambda_T$ is shown in Fig.~\ref{fig:H1_sum_scratch_comb} (b). Comparing the two plots shows that the scratch of width 4 led to much better order near the initial scratch than when the scratch had the larger width 20.4. In fact, there is better order for the scratch of width 4 even when the entire domains are compared; this corresponds to the data points for strip width 400.

We increased the depth of the scratch in the case of a scratch of width 4 by multiplying the first term on the right-hand side of Eq.~(\ref{eq:scratch_init}) by a factor of 10.
Once again, our simulations showed that a ten-fold increase in the template amplitude leads to no reduction in the hexagonal order.

%%%%%%%%%%%%%%%%%%%%%%%%%%%%%%%%%%
\section{Summary}
\label{sec:summary}
Our simulation results show that templating the surface of a binary material prior to ion bombardment can significantly improve the order of nanoscale patterns produced by sufficiently high ion fluences. When the initial wavelengths were approximately one or two times the linearly selected wavelength, the hexagonal and sinusoidal templates both produced dramatically improved global hexagonal order. In particular, the hexagonal templates with initial wavelength approximately double the linearly selected wavelength $\lambda_T$ evolved to a final state which was defect free and perfectly ordered. Impressive improvements in order were also obtained from the scratched templates when the scratch widths were close to or smaller than $\lambda_T$. Moreover, the well-ordered regions of nanodots were centered on the scratches. The results of these simulations with initial scratches demonstrate the potential of using templated samples to produce controllable and localized improvements of the order in nanoscale patterns.

In addition to the positive aspects just discussed, our simulations expose some limitations to the effectiveness of prepatterning. For the hexagonal and sinusoidal templates, we did not observe significantly improved order when the initial wavelength was more than double the linearly selected wavelength. The scratch initial condition, on the other hand, has little effect on the surface at long times if the scratch width is significantly larger than the linearly selected wavelength. Finally, both the sinusoidally-templated and scratched surfaces developed an underlying, long-wavelength rolling topography, which could be problematic in some applications. 

The prepatterns investigated in this paper are not just of academic interest; there are also a practical methods of producing them. Sinusoidal and hexagonal templates could be produced using standard lithographic methods. Scratches could be produced either by dragging an atomic force microscope tip across a sample \cite{Tseng11}, or by scanning a laser or focused ion beam across it. These fabrication techniques would not only produce the desired prepatterns, but could produce them at the length scales that our simulations indicate are needed to observe enhanced ordering.

To understand the pattern formation that is produced by ion bombardment of a binary material,
the coupling between the surface topography and composition must be be taken into account \cite{Bradley10,Shipman11,Bradley11}.  The same is true of two closely related problems:
bombarding an initially elemental material with a beam of metallic ions \cite{Mollick14,Khanbabaee14a,Khanbabaee14b},
and bombarding an initially elemental material with a noble gas ion beam with concurrent deposition of metallic impurities \cite{Bradley11,Bradley12,Bradley13,Ozaydin05,Ozaydin08,Hofsass08,Sanchez-Garcia08,Ozaydin-Ince09,Macko10,Zhang10,Zhou11,Cornejo11,Zhang11,Macko11,Hofsass12,Redondo-Cubero12,Zhang12,Macko12}.
The proposed equations of motion for these three problems have many features in common
\cite{Bradley10,Shipman11,Mollick14,Bradley11b,Bradley12,Bradley13}.
Our finding that templating can lead to improved order in the patterns on binary materials is therefore expected to carry over to the two problems in which metal atoms are implanted in a surface layer.

%%%%%%%%%%%%%%%%%%%%%%%%%%%%%%%%%%%
\section*{Acknowledgments}
DAP thanks Christopher Strickland for porting code from Matlab to Python, and Matthew P. Harrison for helpful discussions. We thank the National Science Foundation for supporting this work through grant DMR-1305449.

\vfill\eject

\end{document}